\documentclass[fleqn,usenatbib]{mnras}

\usepackage{newtxtext,newtxmath}
\usepackage[T1]{fontenc}

\DeclareRobustCommand{\VAN}[3]{#2}
\let\VANthebibliography\thebibliography
\def\thebibliography{\DeclareRobustCommand{\VAN}[3]{##3}\VANthebibliography}

\usepackage{graphicx}	% Including figure files
\usepackage{amsmath}	% Advanced maths commands
\usepackage{caption}
\usepackage{xcolor}
\usepackage{graphbox}

\newcommand{\gs}{\mathrel{\lower0.6ex\hbox{$\buildrel {\textstyle >}
 \over {\scriptstyle \sim}$}}}
\newcommand{\ls}{\mathrel{\lower0.6ex\hbox{$\buildrel {\textstyle <}
 \over {\scriptstyle \sim}$}}}

\newcommand{\lta}{\mathrel{\spose{\lower 3pt\hbox{$\mathchar"218$}}
     \raise 2.0pt\hbox{$\mathchar"13C$}}}
\newcommand{\gta}{\mathrel{\spose{\lower 3pt\hbox{$\mathchar"218$}}
     \raise 2.0pt\hbox{$\mathchar"13E$}}}

\newcommand{\Lya}{Ly$\alpha$}

\newcommand{\Msunpyr}{~\mathrm{M}_{\odot}\,\mathrm{yr}^{-1}\,}

\newcommand{\oii}{\mbox{[O\,{\sc ii]\sc{$\lambda$3727}}}\,}

\newcommand{\oiiia}{\mbox{[O\,{\textsc{iii}}]}\,}
\newcommand{\oiii}{\mbox{[O\,{\sc iii]\sc{$\lambda$5008}}}\,}
\newcommand{\oiiil}{\mbox{[O\,{\sc iii]\sc{$\lambda$4960}}}\,}
\newcommand{\nii}{\mbox{[N\,{\sc ii]\sc{$\lambda$6585}}}}

\newcommand{\siii}{\mbox{[S\,{\sc iii]\sc{$\lambda$9069}}}\,}

\newcommand{\halpha}{\mbox{H\,{\sc$\alpha$}}\,}
\newcommand{\hbeta}{\mbox{H\,{\sc$\beta$}}}
\newcommand{\Pa}{Pa\,$\alpha$\,}
\newcommand{\Pb}{Pa\,$\beta$\,}

\DeclareFontEncoding{LS1}{}{}
\DeclareFontSubstitution{LS1}{stix}{m}{n}
\DeclareSymbolFont{symbols4}{LS1}{stixbb}{m}{it}
\DeclareMathSymbol{\varhexagonblack}{\mathord}{symbols4}{"DD}
\DeclareMathSymbol{\hexagonblack}   {\mathord}{symbols4}{"DE}

\title[The \emph{JWST} Emission Line Survey (JELS)]{The \emph{JWST} Emission Line Survey (JELS): Extending rest-optical narrow-band emission line selection into the Epoch of Reionization}

\author[K. J. Duncan et al.]{K. J. Duncan,$^{1}$\thanks{E-mail: kdun@roe.ac.uk (KJD)}
D. J. McLeod,$^{1}$
P. N. Best,$^{1}$
C. A. Pirie,$^{1}$
M. Clausen,$^{1}$
R. K. Cochrane,$^{1,2,3}$
J. S. Dunlop,$^{1}$\newauthor
S. R. Flury,$^{1}$
J. E. Geach,$^{4}$
N. A. Grogin,$^{5}$
C. L. Hale,$^{6,1}$,
E. Ibar$^{7,8}$,
R. Kondapally,$^{1,9}$
Zefeng Li,$^{7}$\newauthor
J. Matthee,$^{10}$
R. J. McLure,$^{1}$
Luis Ossa-Fuentes$^{7}$,
A. L. Patrick,$^{1}$
Ian Smail,$^{9}$
D. Sobral,$^{11,12}$\newauthor
H. M. O. Stephenson,$^{13}$ 
J. P. Stott,$^{13}$ and 
A. M. Swinbank$^{9}$
\\
$^{1}$Institute for Astronomy, University of Edinburgh, Royal Observatory, Blackford Hill, Edinburgh, EH9 3HJ, UK\\
$^{2}$Department of Astronomy, Columbia University, New York, NY 10027, USA\\
$^{3}$Jodrell Bank Centre for Astrophysics, University of Manchester, Oxford Road, Manchester M13 9PL, UK\\
$^{4}$Centre for Astrophysics Research, School of Physics, Engineering and Computer Science, University of Hertfordshire, Hatfield, UK\\
$^{5}$Space Telescope Science Institute, 3700 San Martin Drive, Baltimore, MD 21218, USA\\
$^{6}$Astrophysics, Department of Physics, University of Oxford, Denys Wilkinson Building, Keble Road, Oxford, OX1 3RH, UK\\
$^{7}$Instituto de F\'isica y Astronom\'ia, Universidad de Valpara\'iso, Avda. Gran Breta\~na 1111, Valpara\'iso, Chile\\
$^{8}$Millennium Nucleus for Galaxies (MINGAL)\\
$^{9}$Centre for Extragalactic Astronomy, Department of Physics, Durham University, South Road, Durham DH1 3LE, UK\\
$^{10}$Institute of Science and Technology Austria (ISTA), Am Campus 1, 3400 Klosterneuburg, Austria\\
$^{11}$Departamento de F\'isica, Faculdade de Ci\`encias, Universidade de Lisboa, Edif\'icio C8, Campo Grande, PT1749-016 Lisbon, Portugal\\
$^{12}$BNP Paribas Corporate \& Institutional Banking, Torre Ocidente Rua Galileu Galilei, 1500-392 Lisbon, Portugal\\
$^{13}$Department of Physics, Lancaster University, Lancaster LA1 4YB, UK\\
}

\date{Accepted 2025 June 27. Received 2025 June 27; in original form 2024 October 14}

\pubyear{2025}

\begin{document}
\label{firstpage}
\pagerange{\pageref{firstpage}--\pageref{lastpage}}
\maketitle

% Abstract of the paper
\begin{abstract}
We present the \emph{JWST} Emission Line Survey (JELS), a \emph{JWST} imaging programme exploiting the wavelength coverage and sensitivity of the Near-Infrared Camera (NIRCam) to extend narrow-band rest-optical emission line selection into the epoch of reionization (EoR) for the first time, and to enable unique studies of the resolved ionised gas morphology in individual galaxies across cosmic history.
The primary JELS observations comprise $\sim4.7\mu$m narrow-band imaging over $\sim63$ arcmin$^{2}$ designed to enable selection of H\,$\alpha$ emitters at $z\sim6.1$ and a host of novel emission-line samples, including [O\,{\textsc{iii}}] ($z\sim8.3$) and Paschen $\alpha/\beta$ ($z\sim1.5/2.8$).
For the F466N/F470N narrow-band observations, the emission-line sensitivities achieved are up to $\sim2\times$ more sensitive than current slitless spectroscopy surveys (5$\sigma$ limits of 0.8--1.2$\times10^{-18}\,\text{erg s}^{-1}\text{cm}^{-2}$), corresponding to unobscured H\,$\alpha$ star-formation rates (SFRs) of 0.9--1.3 $\text{M}_{\odot}\,\text{yr}^{-1}$ at $z\sim6.1$, extending emission-line selections in the EoR to fainter populations.
Simultaneously, JELS also adds F200W broadband and F212N narrow-band imaging (H\,$\alpha$ at $z\sim2.23$) that  probes SFRs $\gtrsim5\times$ fainter than previous ground-based narrow-band studies ($\sim0.2\,\text{M}_{\odot}\,\text{yr}^{-1}$), offering an unprecedented resolved view of star formation at cosmic noon.
We present the detailed JELS survey design, key data processing steps specific to the survey observations, and demonstrate the exceptional data quality and imaging sensitivity achieved.
We then summarise the key scientific goals of JELS, demonstrate the precision and accuracy of the expected redshift and measured emission line recovery through detailed simulations, and present examples of spectroscopically confirmed \halpha and \oiiia emitters discovered by JELS that illustrate the novel parameter space probed.
\end{abstract}

% Select between one and six entries from the list of approved keywords.
% Don't make up new ones.

%%%%%%%%%%%%%%%%%%%%%%%%%%%%%%%%%%%%%%%%%%%%%%%%%%

%%%%%%%%%%%%%%%%% BODY OF PAPER %%%%%%%%%%%%%%%%%%

\section{Introduction}
Since its launch, \emph{JWST} has been delivering on its promise to transform our understanding of the earliest stages of galaxy formation, discovering a wealth of galaxies out to unprecedented redshifts \citep[e.g.][]{robertson2024,Carniani2024}, routinely providing spectroscopic confirmation of galaxies at $z > 10$ \citep[e.g.][]{curtis-lake2023,arrabalharo2023} and discovering new populations of active galactic nuclei \citep[AGN, e.g.][]{Labbe2023,Matthee2023A}, whilst also beginning to reveal the detailed properties of the galaxies which powered the process of cosmic reionization \citep{sanders2023, shapley2023a}.
These early \emph{JWST} results have demonstrated the potential to begin answering some of the key outstanding questions in extra-galactic astronomy. For example, how and when do the first galaxies assemble? How does the chemical enrichment of the Universe proceed \citep{arellano2022, Cameron2023, Curti2023, Isobe2023,topping2024}? When do the first supermassive black holes (SMBHs) form \citep{Labbe2023,larson2023a,maiolino2023,Greene2024,Matthee2024}? Which galaxies drive the process of reionization and what is its detailed topology \citep{tang2023,umeda2024,Mascia2024,Witstok2024}?

However, to date, the majority of $z > 6$ galaxy samples confirmed by the \emph{JWST} Near-Infrared Spectrograph \citep[NIRSpec;][]{Jakobsen2022} have typically been selected on the basis of broadband colours \citep{degraaff2024} or photometric redshift estimates \citep[e.g.][]{bunker2023,hu2024,Maseda2024}.
Such selections, however, can be limited by strong biases, with photometric redshift estimates highly dependent on prior assumptions on the UV continuum slopes and emission line properties \citep{arrabalharo2023, larson2023b}.
To constrain models of galaxy assembly and understand the processes driving early galaxy evolution \citep[][]{somerville2015, yung2019}, it is critical to study stellar mass or star-formation rate (SFR) selected samples of galaxies across cosmic time that are as complete and unbiased as possible (and where any remaining biases can be easily modelled).

The increasing strength of optical emission lines at $z > 2$ has been observed extensively \citep{MarmolQueralto2016, khostovan2016, Reddy2018}. This trend, along with the ubiquity of high equivalent width emission lines in $z > 5$ galaxies \citep[e.g.][]{DeBarros2019} can therefore be used to our advantage, through the efficient selection of galaxies based on their rest-optical emission lines. 
Slitless spectroscopic surveys \citep{Kashino2023,Oesch2023} have already showcased the scientific potential of emission-line-selected samples with the \emph{JWST}. These surveys provide a powerful method to trace the evolution of star-forming galaxies and AGN throughout cosmic history \citep{Matthee2023A, covelopaz2024, Meyer2024}.

Complementary to the grism or slitless spectroscopic approach is the selection of emission-line galaxies using photometric narrow-band observations.
Like slitless spectroscopy, a key advantage of narrow-band selection is that galaxies are selected on the strength of their emission lines, broadly representing a star-formation rate-selected sample. 
When compared to broadband photometric selections, narrow-band surveys also offer the advantage that the robustly selected samples lie within a narrow redshift range while minimising the complex selection effects and biases (e.g. source blending) of slitless samples.
As such, in recent years deep ground-based narrow-band surveys in the optical and near-IR have enabled measurements of the \halpha luminosity function out to $z\sim2.2$ \citep{Geach2008, Sobral2013,Matthee2017}, providing robust measurements of the cosmic star-formation rate density, and enabled detailed studies of the morphology \citep{Sobral2009, Sobral2016}, clustering \citep{Sobral2010, Geach2012, Cochrane2017,Cochrane2018} and environment \citep{Sobral2011} of star-forming galaxies.
Finally, narrow-band selections also offer the advantage of providing a direct resolved view of the ionised gas in individual galaxies. 
Previous narrow-band studies have also provided the samples for high-resolution follow-up observations that enable more detailed morphological, chemical and dynamical studies \citep[e.g.][]{Swinbank2012,Stott2014,molina2019, Cheng2020,Cochrane2021}.

With the inclusion of narrow-band filters on the Near-Infrared Camera \citep[NIRCam;][]{Rieke2005, Rieke2023} from 1.645\micron\, all the way out to 4.7\micron, \emph{JWST} now offers the potential for narrow-band selection of key optical emission lines such as \halpha and \oiiia out into the epoch of reionization (EoR; $z\sim6.1$ and 8.3 respectively).
Simultaneously, the longest wavelength NIRCam narrow-bands can also probe emission lines such as \Pa and \Pb that were previously largely inaccessible in galaxies at the peak of cosmic star-formation history ($1 \lesssim z \lesssim 3$).
The short wavelength ($< 2.5\micron$) narrow-band filters, while not breaking new redshift ground, can exploit \emph{JWST}'s exceptional resolution and sensitivity to constrain the detailed morphology of ionised gas in galaxies at cosmic noon while also probing substantially fainter populations than previously possible from the ground \citep[e.g.][]{Geach2008, Sobral2013, Matthee2017}.

The \emph{JWST} Emission Line Survey (JELS; GO \#2321, PI: Best) is a Cycle 1 NIRCam imaging programme designed to explore this new discovery space using narrow-band observations to detect and study emission-line galaxies across cosmic time.
JELS leverages the extraordinary wavelength coverage, spatial resolution, and sensitivity of \emph{JWST}/NIRCam to extend narrow-band galaxy selections into previously inaccessible observational regimes.
Specifically, using the F466N/F470N filters at $\sim4.7\micron$, the primary goal of JELS is to provide a clean \halpha emission selected sample of galaxies in the EoR ($z=6.1$) that provides complementary constraints on the cosmic star-formation history and whose properties can be characterised and compared against UV-selected samples \citep{Pirie2024}.
Simultaneous 2.12\micron\, (F212N) narrow-band imaging is designed to probe a factor of $\sim5$ deeper than previous ground based studies of \halpha at $z\sim2$ \citep[e.g.][]{Sobral2012}, doing so with sub-kpc resolution to reveal the distribution in star forming galaxies at the peak of cosmic star formation activity in unprecedented detail.
In this paper, we present an overview of JELS; outlining the survey design, the JWST/NIRCam observations, the corresponding data processing and resulting properties of the JELS imaging.
We then summarise the broader scientific goals of the survey, with illustrations of the unique statistical and resolved studies it enables.
Finally, we demonstrate the practical capabilities of the JELS imaging through realistic simulations of $z > 6$ emission-line galaxy populations, as well as presenting illustrative examples of spectroscopically confirmed \halpha and \oiii emitters identified by JELS.

The rest of this paper is set out as follows. In Section~\ref{sec:overview} we present the technical design of the JELS survey, the data reduction process and the photometric properties of the JELS imaging.
In Section~\ref{sec:goals}, we outline the scientific goals of JELS and the expected scientific returns.
In Section~\ref{sec:sims}, we present detailed simulations of the redshift and emission line precision that can be achieved by JELS \halpha and \oiii emission-line selected samples. 
Section~\ref{sec:examples} then presents results from a sample of JELS \halpha and \oiii emission-line candidates confirmed by spectroscopic follow-up observations.
Section~\ref{sec:summary} then summarises our conclusions.
Throughout this paper, all magnitudes are quoted in the AB system \citep{OkeGunn1983} unless otherwise stated.
We also assume a $\Lambda$ Cold Dark Matter ($\Lambda$CDM) cosmology with $H_{0} = 70$ km\,s$^{-1}$\,Mpc$^{-1}$, $\Omega_{m}=0.3$ and $\Omega_{\Lambda}=0.7$.

\section{Survey overview}\label{sec:overview}
The overall observing strategy and survey design of JELS was driven by requirements of the primary science case: firstly, to probe sufficient cosmological volume to detect a statistical sample of $z > 6$ \halpha emitters (a prediction of 40--60 based on conversion from the UV luminosity function of \citeauthor{bouwens2015}~\citeyear{bouwens2015} and assuming UV to \halpha conversion as presented in \citeauthor{Hao2011}~\citeyear{Hao2011}), and secondly to probe significantly below the break of the luminosity function for \halpha emitters at $z > 6$, thus identifying `typical' star-forming galaxies comparable to those selected in Lyman-break selected samples but with a highly complementary selection function.

While designed to be scientifically viable with only existing legacy multi-wavelength observations \citep[e.g. CANDELS COSMOS;][]{Grogin2011,Koekemoer2011}, JELS was designed in tandem with the \emph{JWST} Cycle 1 Guest Observer Treasury Program `Public Release Imaging For Extragalactic Research' Survey (PRIMER; GO \#1837, Dunlop et al., \emph{in prep}). 
The complementary broadband observations across the full NIRCam wavelength range from PRIMER improve the reliability of narrow-band excess selection at $\sim 4.7\micron$ (see below), and are critical for panchromatic spectral energy distribution modelling and robust emission-line disambiguation for line emitters selected at both 2.12\micron\, and $\sim4.7$\micron.
An additional survey design criteria was therefore for the narrow-band imaging to be located entirely within both the pre-existing deep optical and contemporaneous PRIMER NIRCam observations.

\subsection{Survey design}\label{sec:survey-design}
\begin{figure*}
\begin{minipage}{\textwidth}
  % \centering
  % \includegraphics[align=c,height=1.25in]{fig1.pdf}
  % \hspace*{.2in}
  % \includegraphics[align=c,height=1in]{fig2.pdf}
    \centering
    \includegraphics[align=c, width=0.48\columnwidth]{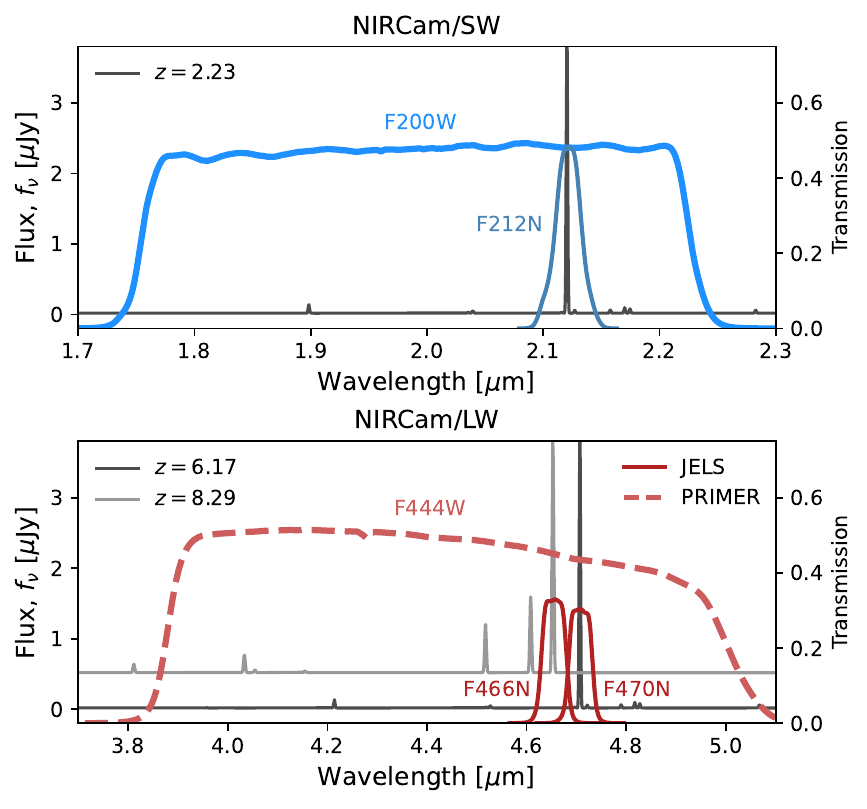}
    \hspace*{.1in}
    \includegraphics[align=c,width=0.47\columnwidth]{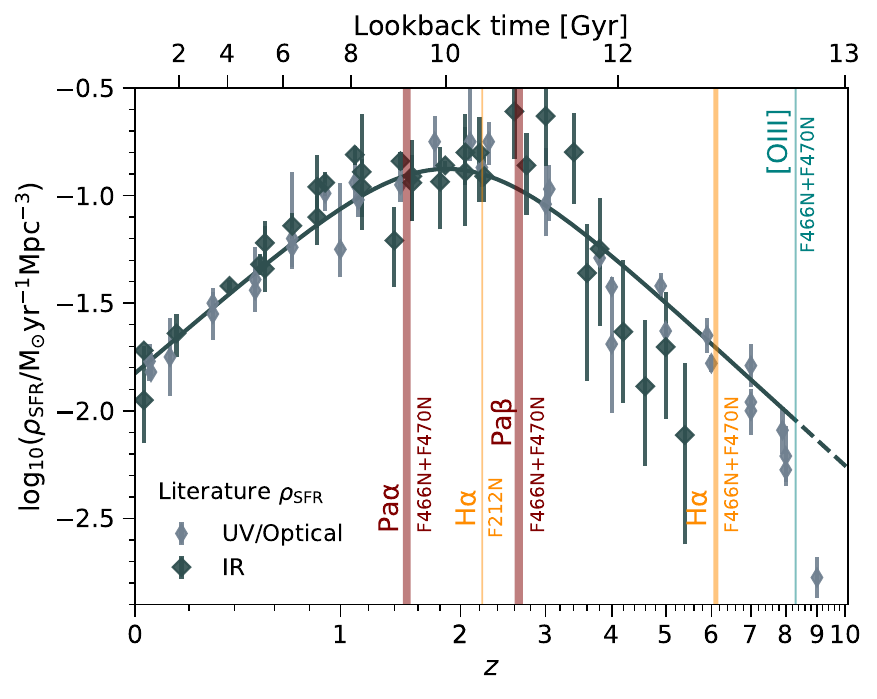}
    \caption{Left: Illustration of the JELS filter-set in the NIRCam/SW (upper panel) and LW bands (lower panel). Shown for reference is an illustrative star-forming galaxy spectral energy distribution redshifted to key redshifts where the narrow-band filters probe \halpha ($z=2.23$ for F212N, $z=6.17$ for F470N) and \oiii ($z=8.29$ for F466N). The continuum flux density values are normalised arbitrarily for visualisation purposes. Right: Key emission lines and SFR-indicators probed by the JELS narrow-band filters in the context of the cosmic SFR density evolution. Literature UV/Optical (narrow diamonds) and infrared (wide diamonds) SFR density measurements are shown from the compilation by \citet[][see references therein]{Madau2014} and illustrative additional measurements at $z > 3$ from recent studies in the rest-UV \citep{Bouwens2022} and far-IR \citep[][for 850\micron\, sources brighter than $>0.2\text{mJy}$]{dudzeviciuite2020}.}
    \label{fig:filter_example}
\end{minipage}
\end{figure*}

In the NIRCam long-wavelength (LW) channels, JELS employs the closely-separated F466N and F470N narrow-band filters, with pivot wavelengths of 4.654 and 4.707\micron\, respectively (and effective filter widths, $W_{\text{eff}}$, of 0.054 and 0.051\micron).
Either through difference imaging between the adjacent F466N/F470N bands, or narrow-band excess selection with respect to the F444W broadband imaging provided by the PRIMER COSMOS observations, line emitters can be selected in both filters and enable selection of \halpha emitters in two overlapping volumes centred at $z\sim6.09$ and $z\sim6.17$ (see Fig.~\ref{fig:filter_example}).

Simultaneously, in the NIRCam short-wavelength (SW) channels, JELS employs the F212N narrow ($\lambda_{\text{pivot}}=2.12\micron\,$, $W_{\text{eff}}=0.027\micron$) and F200W broadband ($\lambda_{\text{pivot}}=1.99\micron\,$, $W_{\text{eff}}=0.419\micron$) filters with a traditional narrow--broadband approach (Fig.~\ref{fig:filter_example}). 
The F212N filter probes \halpha at $z=2.23$, corresponding to the peak epoch of cosmic star formation, significantly extending the luminosity range of existing ground-based studies at this key redshift while also enabling resolved studies of the ionised gas at sub-kpc resolution.
Besides ensuring matched broadband coverage for F212N narrow-band selection, the F200W broadband imaging also adds $2-3\times$ the PRIMER exposure times within the centre of the field, providing significant gains in sensitivity for broader galaxy evolution studies in this critical legacy field.

% \begin{figure}
%     \centering
%     \caption{}
%     \label{fig:csfrd}
% \end{figure}

JELS uses a $3\times3$ mosaic strategy with 5 per cent overlap in each row and 57 per cent overlap between each column. 
We also adopt the standard 3-point intra-module dithering at each location to fill in short-band intra-chip gaps and account for both bad pixels and cosmic rays, with sub-pixel shifts at each primary dither position.
The full JELS mosaic provides contiguous coverage over a total area of $\sim63\, \text{arcmin}^{2}$.
Combining the wavelength coverage of both F466N and F470N, this area corresponds to a \halpha selection volume of $\sim2.4\times10^{4}~\text{Mpc}^{3}$ (at $z\sim6.1$).
For F212N, the single narrower filter results in an approximate \halpha selection comoving volume of $\sim0.9\times10^{4}~\text{Mpc}^{3}$ (at $z\sim2.2$).

For the NIRCam SW and LW filter pairs, all observations use the MEDIUM-10 readout strategy, with 6 total integrations/dithers at each mosaic position.
The F212N and F470N(+F444W)\footnote{Both F466N and F470N are observed with F444W as a blocking filter. We use `(+F444W)' here to denote that.\label{ft:blocking}} filter combination was observed with 10 groups per integration, with this observing setup resulting in an on-sky integration time of 6\,313s over the full mosaic.
Over the central $\sim40$ per cent of the mosaic that is imaged twice, the total exposure time reaches 12\,626s.
The F200W and F466N(+F444W)\footref{ft:blocking} filter combination was observed with 9 groups per integration with the same number of total integrations and dithers, yielding a total on-sky integration time of 5\,669s over the full mosaic (11\,338s over the central 40 per cent).
Note that the difference in groups per integration between F466N and F470N observations (9 versus 10) is a result of changes to the readout strategy from the initial proposal (DEEP-10 to MEDIUM-10) to reduce the impact of cosmic rays, and the additional overheads that resulted in.
The F212N/F470N pairing was prioritised to maximise the sensitivity in F212N. 

\subsection{Observations and image reduction}
The initial JELS observations were acquired over the period of May 6 to May 27 2023.
Of these observations, 13 of 18 visits were observed without issue. 
In five of the eighteen total visits of the JELS observations (one F200W+F466N, four F212N+F470N), the NIRCam imaging was subject to unusually bright scattered light, or `wisps', that resulted in features in \emph{both} the SW and LW detectors.\footnote{See e.g. the JWST Known Issues: \href{https://jwst-docs.stsci.edu/known-issues-with-jwst-data/nircam-known-issues/nircam-scattered-light-artifacts\#NIRCamScatteredLightArtifacts-wisps}{https://jwst-docs.stsci.edu/known-issues-with-jwst-data/nircam-known-issues/nircam-scattered-light-artifacts\#NIRCamScatteredLightArtifacts-wisps}\label{ft:scat_light}}
Initial tests demonstrated that the LW wisps remained spatially invariant within the field of view and have approximately constant amplitude within a visit and could therefore be effectively removed as set out below.

However, unlike most common wisps, the scattered light in the impacted JELS SW images is both extremely bright and shows notable small-scale variation in morphology between exposures within a given visit.
With the corresponding LW images also impacted, standard approaches to mitigate the scattered light are not viable \citep[see e.g.][]{Robotham2023}.
For the impacted F200W+F466N visit, the availability of the separate PRIMER F200W observations and the limited impact on F466N mean that the overall impact is negligible.
However, for the four F212N+F470N visits where the F212N exposures are not viable for the proposed scientific goals, these specific pointings were re-scheduled and successfully observed over the period of November 23 to November 24 2024.
% The F212N analysis (and to a lesser degree F470N) presented below therefore does not represent the final expected JELS image quality and sensitivity.

In the following sections we describe the current version of the data reduction and the resultant imaging properties from this full dataset (v1.0). Other early JELS analysis \citep{Pirie2024} were based on an earlier data reduction, in which the re-observed frames were not included (v0.8). For completeness, in Appendix~\ref{sec:app-versions} we describe the differences between the v0.8 and v1.0 datasets.

\begin{figure*}
\centering
\includegraphics[trim={0.4cm 0 0 0}, width=0.59\textwidth]{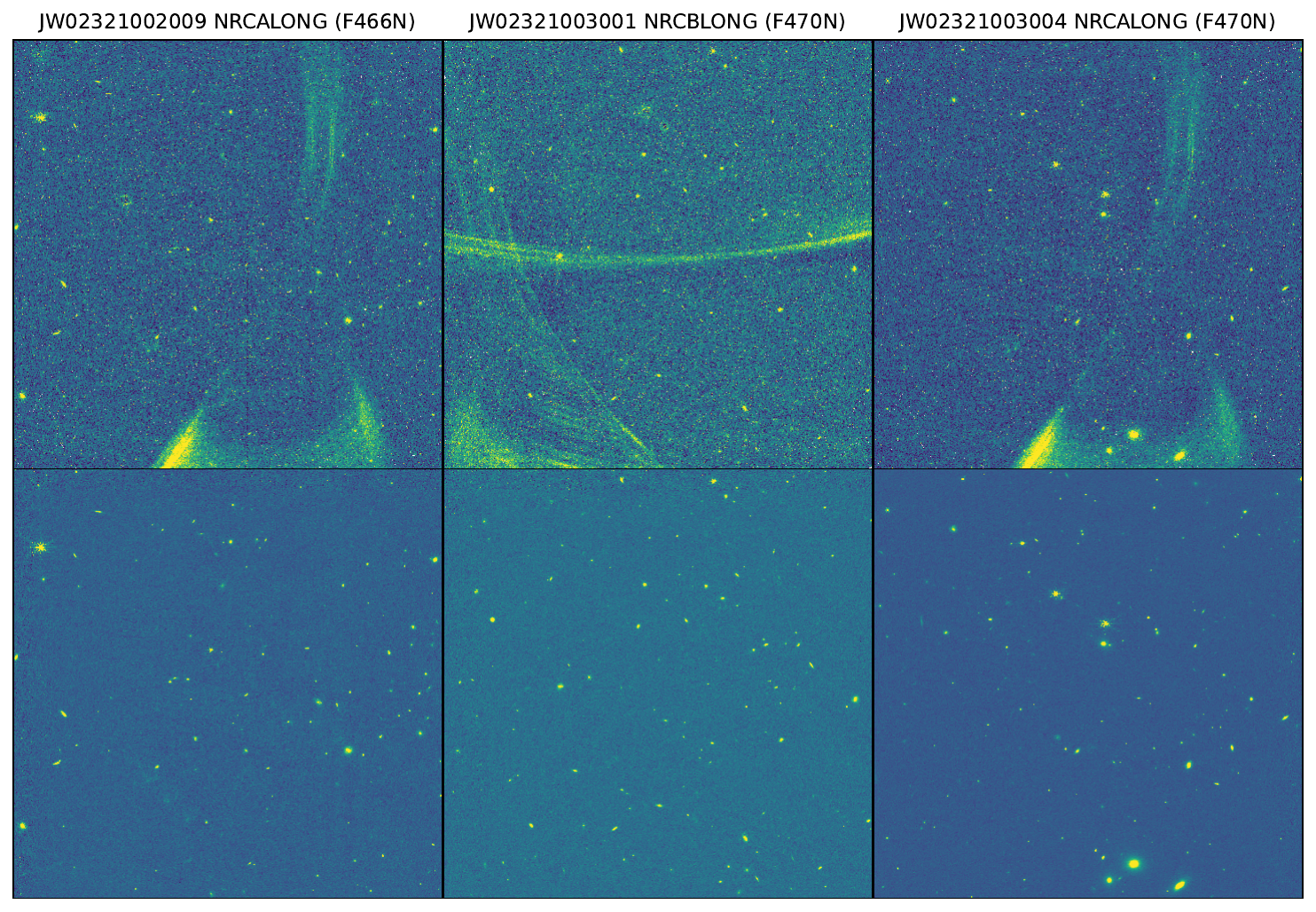}
    \includegraphics[width=0.39\textwidth]{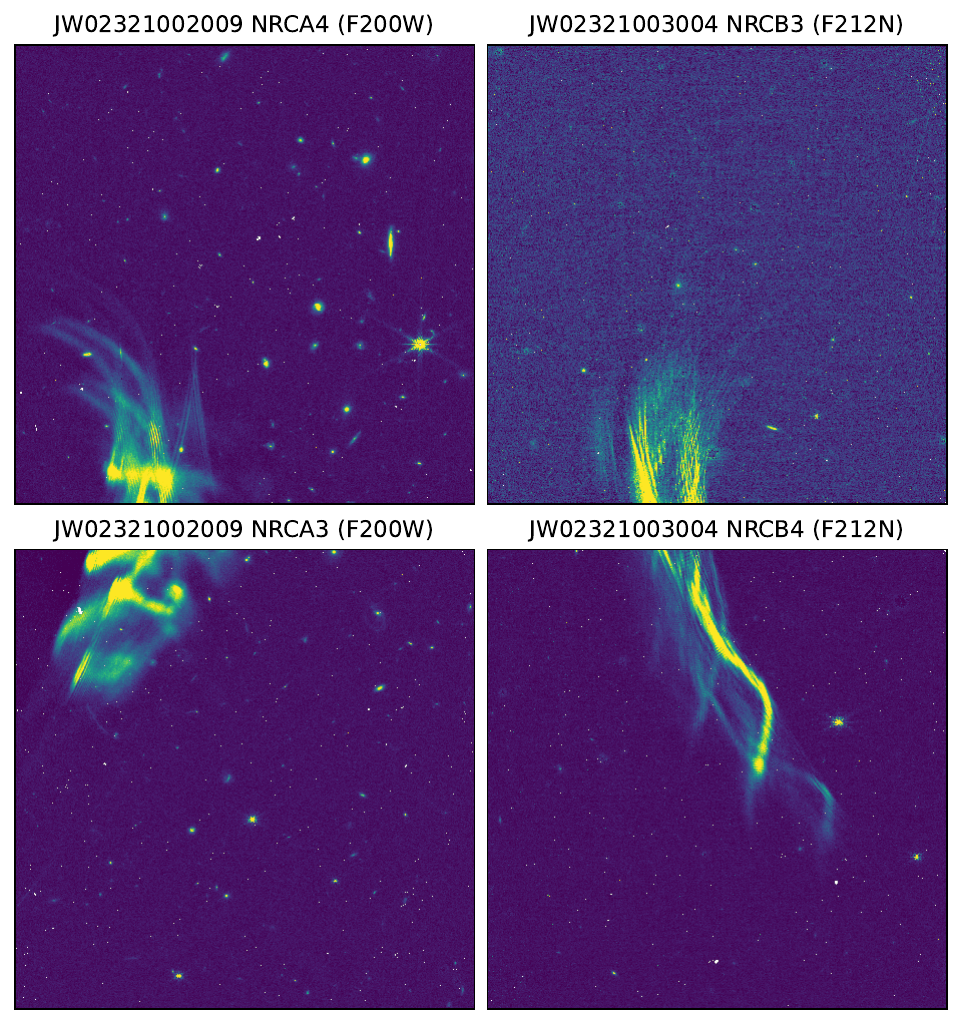}
	\caption{Left: Illustration of the transient scattered light and associated removal for the NIRCam long wavelength imaging. The top row shows reduced individual frames for a single exposure in three of the F466N/F470N visits impacted by scattered light, labelled with the corresponding observation ID, NIRCam module and filter. The bottom row shows the corresponding region of sky from the final mosaic after subtraction of the scattered light in individual frames and all overlapping dithers or mosaic visits have been combined. Each region (i.e. column) is shown with the same colour scale, with a linear stretch between 2 and 99.5 per cent of the individual impacted visit. Right: Illustration of the more extreme scattered light impacting the F200W and F212N observations.}
    \label{fig:scattered_light}
\end{figure*}

\subsubsection{Imaging pipeline}
To ensure consistency with the key ancillary imaging in the field, all JELS imaging is
processed through the Primer Enhanced NIRCam Image Processing Library (PENCIL; Magee et al., in prep) software, an enhanced version of the \emph{JWST} pipeline (version 1.13.4) and using the \texttt{jwst\_1303.pmap} Calibration Reference Data System (CRDS) file.
This PENCIL pipeline includes additional routines for the removal of snowballs and standard wisp artefacts, correction for $1/f$ noise striping as well as background subtraction.
The astrometry of the reduced images is tied to the Gaia DR3 \citep{Gaia2022} reference frame and stacked to a consistent pixel scale ($0.03 \,\textrm{arcsec}\times0.03\,\textrm{arcsec}$).
Prior to the construction of the final mosaics, we also take additional steps to mitigate the impact of the non-standard wisp artefacts in the five impacted JELS visits. 

\subsubsection{Scattered light removal}
For the affected LW images that drive the primary JELS science aims, subtraction of the wisps is possible, but requires a specific approach tuned to these observations.
In NIRCam observations impacted by the typical wisp scattered light patterns, the robust prior positions (and spatial extent) of real sources in the affected SW modules can be obtained from the corresponding unaffected LW module image \citep{Robotham2023}.
For the severe wisps in JELS, this is not possible since the LW frames themselves are affected.
Templates for the LW scattered light therefore must be derived directly from the observed data and make use of the fact that our observing strategy provides 3 intra-module dithers at each pointing.

The impacted frames are first processed through the initial PENCIL data reduction stages as above.
A scattered light template for a given set of exposures is then generated as follows:
\begin{enumerate}
    \item Compact sources within the image are identified, with source detection employing a small locally varying background ($25\times25$ pixel box-size) to select genuine sources within regions of diffuse scattered light. In each exposure, the compact sources are then masked with additional dilation around each source, consisting of two iterations using a binary dilation structure with connectivity equals one (i.e. no diagonal elements are neighbours).
    \item The six frames are median stacked in the image plane to produce a wisp template for the frame. The median stacked template is then convolved with a Gaussian smoothing kernel with a full-width half-maximum (FWHM) of 5 pixels to reduce small-scale noise (with interpolation for any remaining masked pixels).
    \item A template mask is then generated by smoothing the scattered light template with a larger Gaussian kernel (FWHM $=25$ pixels). Regions of the smoothed template image with flux below $0.5\times\sigma_{\textup{RMS}}$ are set to zero.
\end{enumerate}
The final step is included to avoid the addition of noise from the generated template to pixels that are largely unaffected by the scattered light. 
By construction however, this may leave some areas of fainter diffuse scattered light present within individual frames prior to the final mosaicing and combination of all frames covering the region.
We mitigate the impact of this low-level scattered light outside the masked regions in the following way. First we mask sources with a dilated segmentation map (size=5, increasing to size=20 for the most extended sources). We then make a median stack of the three masked, odd valued dithers and subtracted this from the even valued dithers. The procedure is then applied for subtracting even valued dithers from odd, before the final processed frames were included in the mosaicing.
Fig.~\ref{fig:scattered_light} illustrates the effectiveness of the LW wisp removal on one of the impacted JELS visits.

Due to the extreme brightness and significant exposure-to-exposure variation in the spatial distribution within the affected SW frames, subtraction of the scattered light in SW is not feasible.
However, even for the most severely impacted NIRCam SW modules, the majority of pixels in each frame remain perfectly usable for scientific analysis (see right panels of Fig.~\ref{fig:scattered_light}). 
For the JELS SW data reduction, we therefore follow a similar procedure as above, generating an average scattered light template for each affected module.
The scattered light mask generated from template generation procedure outlined above is instead then used to mask the most severe areas of scattered light before the subsequent mosaicing.

Finally, once scattered light has been either subtracted or masked, the JELS imaging is then stacked onto the same 0.03 arcsec pixel grid as the PRIMER COSMOS imaging \citep[see e.g.][]{Donnan2024} that is tied to the same astrometric reference.
In the regions impacted by severe scattered light, the final SW mosaics and the F470N LW mosaic combine both the initial observations (with the severe scattered light treated following the procedure outlined above) and any additional data from the respective repeat observations. 

\subsection{JELS Data Properties}\label{sec:properties}
\begin{figure*}
    \centering
    \includegraphics[width=0.99\textwidth]{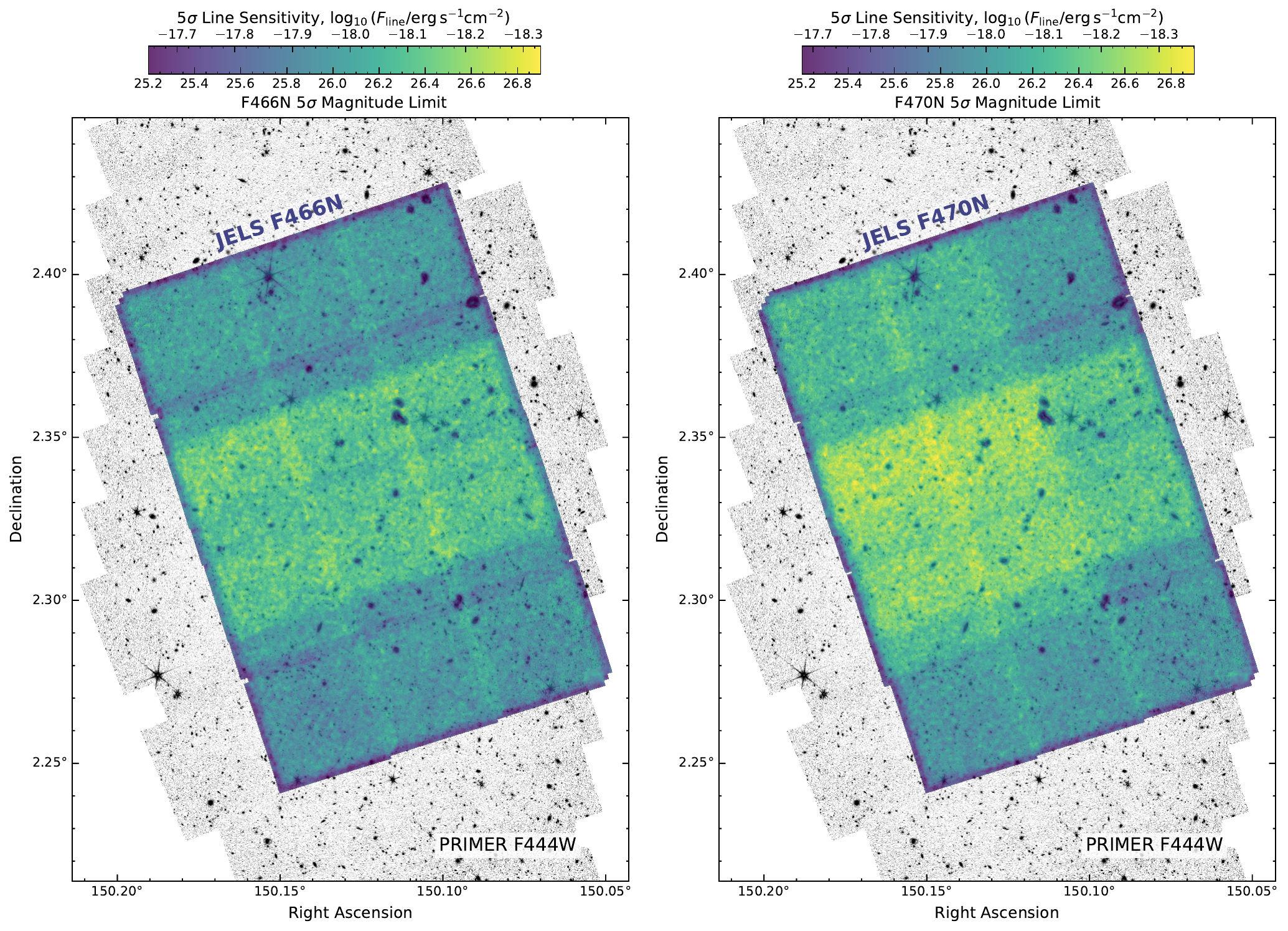}
    \caption{Illustration of the layout and sensitivity of the JELS long-wavelength narrow-band imaging within the PRIMER COSMOS field. The background image shows the PRIMER F444W broadband image with an illustrative image stretch. The 5$\sigma$ limiting magnitude and corresponding maximum emission line sensitivity for the JELS F466N (left) and F470N (right) narrow-band imaging are shown by the colour scale, where the local noise at a given pixel is estimated from the nearest 200 0.3 arcsec  apertures in empty sky regions. The central $\sim 1/3^{\text{rd}}$ of the JELS field that has double the exposure time is immediately apparent in both mosaics. Similarly, the overall gain in sensitivity from repeats of observations most severely impacted by scattered light can be seen in the north east quadrant of the F470N depth map.}
    \label{fig:field_layout}
\end{figure*}

Fig.~\ref{fig:field_layout} illustrates the relative position of the JELS imaging in the context of the PRIMER COSMOS field for the final observed position angles.
With the exception of a small region in the south west of the field (the majority of which is covered by two or fewer dither positions), the JELS LW narrow-band imaging is fully contained within the NIRCam F444W broadband imaging necessary for optimal F466N/F470N emission-line selection.
We estimate the local noise in the JELS imaging by placing down 0.3 arcsec diameter apertures in empty regions of the image, with the 1$\sigma$ noise at a given sky position based on the scatter in the nearest 200 apertures.
For a given limiting flux density in $f_{\nu}$, we then estimate the corresponding maximum emission line sensitivity, $F_{\text{line}} (\text{erg\,s}^{-1} \text{cm}^{-2})$, assuming the flux density in the narrow-band is dominated by the emission line (i.e. $\sim0$ continuum) or the uncertainty on the continuum estimate from the underlying broadband (F444W for F466N/F470N, F200W for F212N) is negligible. 
Note, this conversion also assumes the emission line is centred at the pivot wavelength, $\lambda_{\text{pivot}}$, of the narrow-band filter.
Finally, the 0.3 arcsec aperture line flux limits are corrected to total fluxes for an assumed point-source based on the corresponding fraction of encircled energy for the respective point-spread function (PSF).
The resulting distribution of limiting magnitudes and line fluxes for the F466N imaging is shown by the colour scale in Fig.~\ref{fig:field_layout} with the additional depth achieved in the central region of the field clearly visible.

\begin{figure}
    \centering
    \includegraphics[width=0.98\columnwidth]{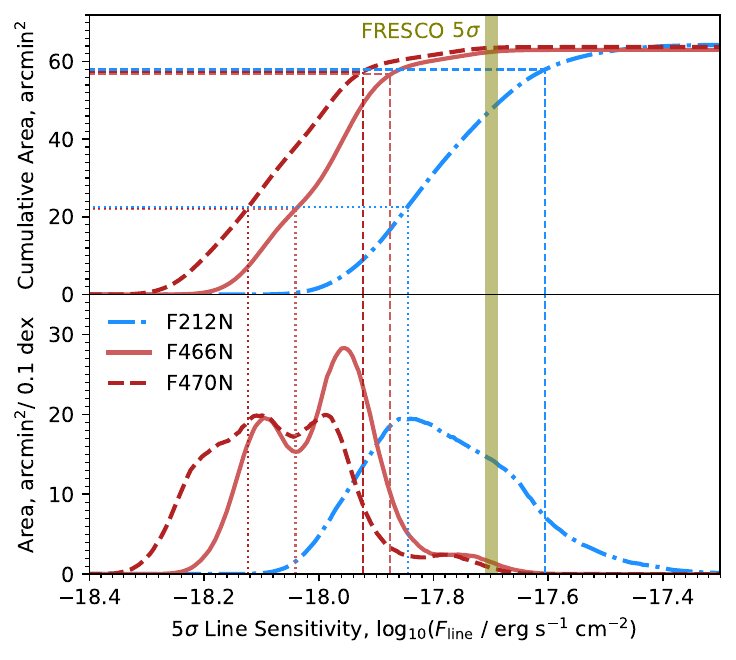}
    \caption{Emission line sensitivity of the JELS narrow-bands as a function of area covered, both cumulative area (top) and area per limiting sensitivity (bottom). The cumulative area corresponding to the deepest 35 per cent ($\sim22$ arcmin$^{2}$; i.e. the central region with double exposure) and 90 per cent ($\sim57$ arcmin$^{2}$) of the full JELS area are marked with thin horizontal dotted/dashed lines respectively, with vertical lines where they intersect the sensitivity curve. Exact values are tabulated in Table~\ref{tab:depths} alongside the corresponding \halpha luminosity limit. Also plotted for reference is  the average 5$\sigma$ limiting line sensitivity for \halpha sources at $4.9 < z < 6.6$ in the FRESCO slitless spectroscopy survey \citep[thick vertical line;][]{Oesch2023}.}
    \label{fig:sensitivity_curves}
\end{figure}

\subsubsection{F466N and F470N Narrow-band Sensitivity}\label{sec:lw_props}
To summarise the distribution of limiting line fluxes for the JELS narrow-band mosaics, in  Fig.~\ref{fig:sensitivity_curves} we present the estimated line flux sensitivity as a function of both cumulative area (upper panel) and area per limiting flux ($\log_{10}(F_{\text{line}}/\text{erg\,s}^{-1} \text{cm}^{-2}$); lower panel).
As designed, the F466N and F470N imaging reach broadly comparable depths, but with F470N reaching higher sensitivity over the full field due to the additional group per integration and the additional repeat observations also increasing the maximum sensitivity reached in the deepest part of the field.
For both F466N and F470N, the additional depth of the central region is immediately apparent (e.g. lower panel of Fig.~\ref{fig:sensitivity_curves}).

To summarise the sensitivities reached in the central and overall field areas, Table~\ref{tab:depths} presents the limiting line flux sensitivities reached for the deepest 35 and 90 per cent of the field (see also the vertical dotted and dashed lines in Fig.~\ref{fig:sensitivity_curves}) in each of the JELS narrow-bands.
Also shown in Table~\ref{tab:depths} is the limiting \emph{observed} \halpha luminosity, $L_{\text{H}\alpha}$, corresponding to the flux limit for the specified redshifts.
The conversion to \halpha luminosity includes a correction for the typical expected contribution from \nii, assuming $\log_{10}(\nii/\halpha) = -1.31$ as measured for high-$z$ galaxies in early \emph{JWST} samples \citep[][which leads to a $0.021\,\text{dex}$ correction to $L_{\text{H}\alpha}$]{shapley2023a}.
In the remainder of the manuscript, all quoted $L_{\text{H}\alpha}$ at $z\sim6$ are corrected for $\nii$ contribution based on this assumption unless explicitly specified (e.g. $L_{\text{H}\alpha+[N\,{\textsc{ii}]}}$).

The measured sensitivity presented in Fig.~\ref{fig:sensitivity_curves} (and Table~\ref{tab:depths}) compares favourably to the pre-launch predicted sensitivity of $2.2\times10^{-18} \,\text{erg s}^{-1} \text{cm}^{-2}$ within a 0.4 arcsec aperture (for an assumed characteristic radius of 0.15 arcsec and S\'{e}rsic $n=1.2$ light-profile) for the areas with $\sim6\,000$s total exposure, extending down to $1.5\times10^{-18} \,\text{erg s}^{-1} \text{cm}^{-2}$ over the central region with double the exposure time.
Although formally not the exact same metric, the measured luminosity limits are between $\sim0.2-0.3\,\text{dex}$ more sensitive than pre-launch predictions, in line with in-flight performance for NIRCam photometry seen more widely in medium and broadband imaging \citep{rigby2023}.

The limiting $L_{\text{H}\alpha}$ corresponds to \emph{unobscured} \halpha star-formation rates of 1.9--3.2 $\text{M}_{\odot}\,\text{yr}^{-1}$ based on low-redshift calibrations \citep{Hao2011}, or 0.9--1.3 $\text{M}_{\odot}\,\text{yr}^{-1}$ for SFR calibrations more appropriate for younger stellar populations at higher redshift \citep[$\log_{10}(L_{\text{H}\alpha}/\text{M}_{\odot}\,\text{yr}^{-1}) = -41.64$;][]{Theios2019}.
We note that the intrinsic \halpha luminosity distribution probed, once corrected for dust attenuation, will naturally be brighter. For example, \citet{covelopaz2024} measure an average extinction of $A_{\text{H}\,\alpha}=0.47$ for $4 < z < 6.5$ \halpha emitters.

Given the $\sim4.7\mu m$ sensitivities achieved over the survey area of $\sim\,63\, \text{arcmin}^{2}$, the JELS narrow-band survey probes a complementary parameter space to the "First Reionization Epoch Spectroscopically Complete Observations" \citep[FRESCO; GO \#1895,][]{Oesch2023} slitless spectroscopy survey, which reaches a 5$\sigma$ line sensitivity at $\sim 4-5\micron$ of $2\times10^{-18} \,\text{erg s}^{-1} \text{cm}^{-2}$ \citep[Fig.~\ref{fig:sensitivity_curves}, see also][]{covelopaz2024}. 
Covering two $\sim62\,\text{arcmin}^{2}$ fields (of which $4.4-5\micron$ is visible over $\sim73$ per cent), the wavelength coverage of FRESCO allows un-targeted detection of emission-line sources over a wider redshift range and hence survey volume (plus simultaneous spectroscopic confirmation).
While the robust identification of single lines like \halpha may be challenging \citep{Baronchelli2020}, continuum subtracted detection techniques have proven to be highly effective \citep[see e.g.][in addition to the citations above]{Helton2024,Meyer2024} for grism surveys.
Although probing a more limited redshift range, the JELS narrow-band is sensitive to fainter line fluxes at $\sim4.7\micron$ over the full mosaic, reaching up to $\sim2\times$ fainter in the central region, with no losses due to dispersion off the detector and reduced blending or contamination from bright sources.
As such, JELS is able to robustly detect a sample of \halpha emitters at $z > 6$ comparable to both FRESCO fields combined, despite the limited volume ($z\sim6.1$ vs $6 < z < 6.6$, see Section~\ref{sec:goals-halpha} below).
Additionally, as further discussed in Section~\ref{sec:morph}, the narrow-band imaging retains the full two-dimensional morphological information, providing information on the spatially resolved ionised gas structures of individual galaxies \citep[cf. the ensemble structural information robustly measurable in slitless spectroscopy;][]{Matharu2024}.
The two approaches are therefore highly complementary, and combined probe a broader dynamic range of emission line galaxies in the EoR.

\subsubsection{F212N Narrow-band Sensitivity}
\begin{figure*}
    \centering
    \includegraphics[width=0.99\textwidth]{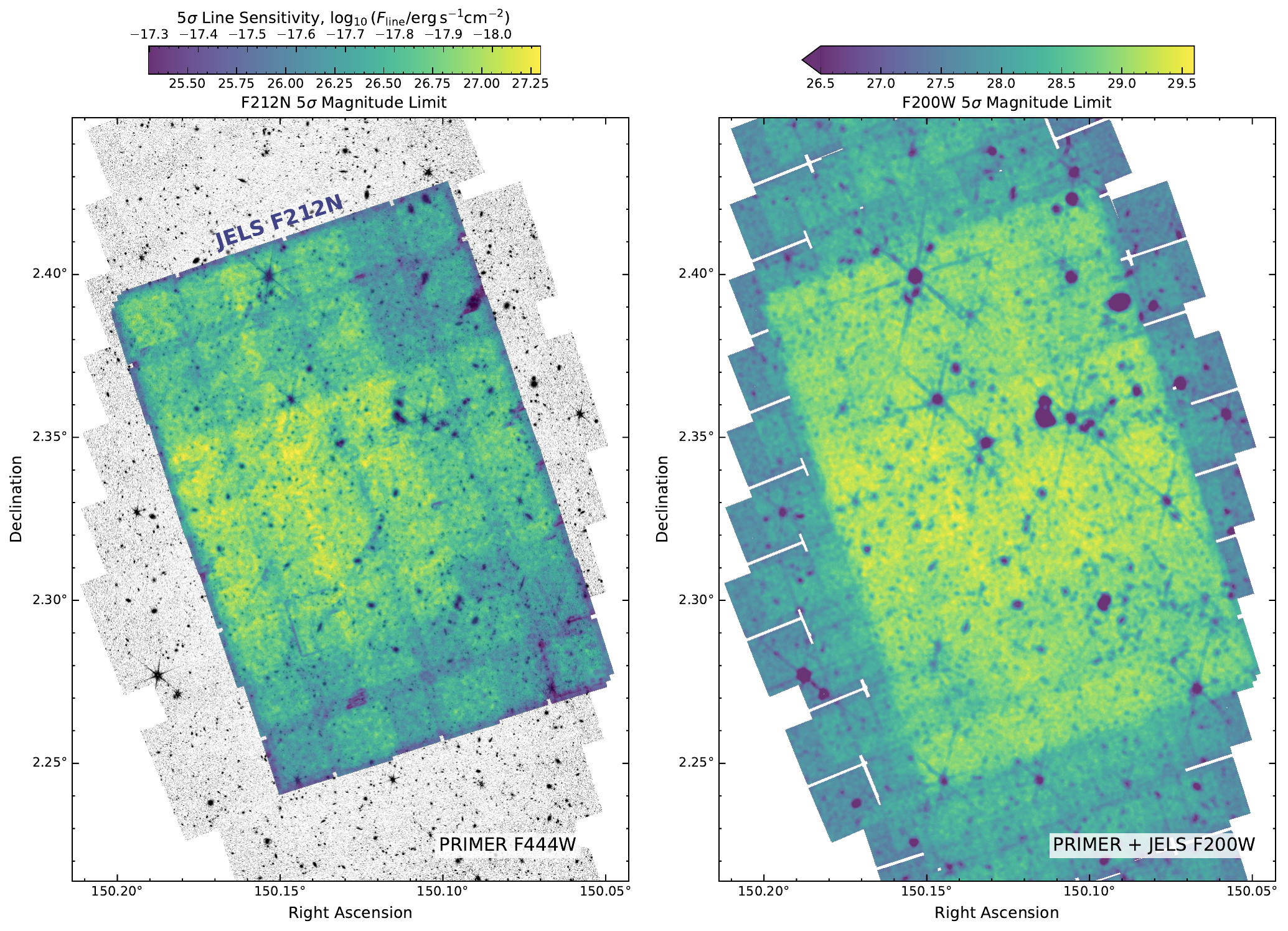}
    \caption{Illustration of sensitivity of the combined PRIMER and JELS F200W imaging in the COSMOS field. The 5$\sigma$ limiting magnitude and corresponding maximum emission line sensitivity are shown by the colour scale, where the local noise at a given pixel is estimated from the nearest 200 0.3 arcsec apertures in empty sky regions and corrected to total fluxes given the fraction of encircled energy for the F200W PSF. As for F470N, the overall gain in sensitivity from repeats of observations impacted by scattered light can be seen in the north east quadrant of the F212N depth map.}
    \label{fig:f200w_depths}
\end{figure*}

For the F212N narrow-band, the mosaic reaches a $5\sigma$ limiting line flux of $2.5\times10^{-18} \text{erg s}^{-1} \text{cm}^{-2}$ for the deepest 90 per cent of the field (see Table~\ref{tab:depths}).
This depth corresponds to $\sim0.7-0.8\,\text{dex}$ fainter \halpha luminosities probed at $z\sim2.2$ than the LW narrow-bands probe at $z\sim6.1$, reaching $L_{\text{H}\alpha} > 10^{40.99} \text{erg s}^{-1}$ ($10^{40.75} \text{erg s}^{-1}$ in the doubly imaged area).
For the same SFR calibration assumed above \citep{Theios2019}, these limits correspond to unobscured \halpha SFRs of 0.13-0.22 $\text{M}_{\odot}\,\text{yr}^{-1}$.
We note that in Fig.~\ref{fig:sensitivity_curves}, the bi-modality corresponding to the central region is less clearly defined for the F212N filter due to the variation in depths from mosaicing of the smaller SW modules and the increased impact of bright stars in the field (c.f the extent of diffraction spikes in Fig~\ref{fig:f200w_depths} relative to Fig.~\ref{fig:field_layout}).
Nevertheless, the JELS imaging still reaches line sensitivities $\sim5\times$ more sensitive than the previous state-of-the-art over the full field and almost an order of magnitude more sensitive in the central region \citep[cf. $\sim10^{41.7} \text{erg s}^{-1}$;][]{Geach2008, Hayes2010, Sobral2013}.

\begin{table}
    \centering
    \caption{Limiting emission line flux and corresponding observed $L_{\text{H}\alpha}$ sensitivity (corrected for \nii\, contribution) reached for the three JELS narrow-band images shown in Fig.~\ref{fig:sensitivity_curves}, with 5$\sigma$ point-source limits corresponding to the deepest 35 per cent of the field ($\sim22$ arcmin$^{2}$; i.e. the central region with double exposure) and the deepest 90 per cent ($\sim57$ arcmin$^{2}$). As noted in the main text, intrinsic $L_{\text{H}\alpha}$ limits accounting for dust attenuation will be higher than the quoted values.}
    \label{tab:depths}
    \begin{tabular}{ccccc}
    \hline
    Filter ($z_{\text{H}\alpha}$) & \multicolumn{2}{c}{5$\sigma$ Flux Limit} & \multicolumn{2}{c}{5$\sigma$  Luminosity Limit} \\
     & \multicolumn{2}{c}{$\log_{10}(F_{\text{line}}/ \text{erg\,s}^{-1} \text{cm}^{-2})$} & \multicolumn{2}{c}{$\log_{10}(L_{\text{H}\alpha}/ \text{erg\,s}^{-1})$} \\ \hline
     & \multicolumn{4}{c}{Percentage of the JELS coverage} \\
     & 35\% & 90\% & 35\% & 90\% \\
    \hline
F212N (2.23) & $-17.84$ & $-17.60$ & 40.75 & 40.99 \\
F466N (6.09) & $-18.04$ & $-17.88$ & 41.60 & 41.76  \\
F470N (6.17) & $-18.12$ & $-17.92$ & 41.53 & 41.73 \\
    \hline
    \end{tabular}
\end{table}

\subsubsection{F200W Broadband Sensitivity}
Finally, in Fig.~\ref{fig:f200w_depths}, we illustrate the additional F200W broadband depth gained in the centre of the PRIMER COSMOS field from the addition of the JELS imaging. 
The F200W mosaic in COSMOS using only PRIMER imaging reaches an average global 5$\sigma$ depth of $m_{\text{F200W}} = 28.3$ over the best 63 arcmin$^{2}$ of the field when the depths are calculated using the same 0.3 arcsec apertures corrected to total flux (we note that the different depths quoted in \citeauthor{Donnan2024}~\citeyear{Donnan2024} are calculated from PSF homogenised images).
With the JELS F200W imaging included, the equivalent deepest area within the field reaches a median depth of $m_{\text{F200W}} = 28.7$.
We note that in Fig.~\ref{fig:f200w_depths}, the impact and extent of bright stars in the field is visibly worse than seen in any of the narrow-band images.
We attribute this difference primarily to the significantly increased continuum sensitivity of the F200W image and the resulting increased dynamic range that makes the presence of extended diffraction spikes more visible.
The full extent of the brighter stars is already visible in just the PRIMER imaging alone, so other broadband filters will be affected to a similar degree.
Additionally, unlike the diffraction patterns for the narrow-band filters (that are somewhat stippled due to the narrow wavelength coverage), the F200W diffraction spikes are continuous and therefore contribute more to the local noise in these regions. 
Given the importance of the ancillary data to the narrow-band selection, bright star masks that fully cover the effected regions will need to be incorporated in subsequent scientific analysis (as is standard in the literature).

\section{Scientific goals}\label{sec:goals}
\subsection{Global line emitter properties}
\subsubsection{A census of \halpha emitters at $z\sim6.1$}\label{sec:goals-halpha}
The primary goal of the JELS survey is to carry out the first narrow-band \halpha survey at $z > 6$, obtaining a sample of $\gtrsim40$ \halpha emitters \citep{Pirie2024}. 
The JELS \halpha sample is designed to provide tight constraints on the faint end of the \halpha luminosity function at $z\sim 6$, with sufficient accuracy to constrain the space density of \halpha emitters at these luminosities to within 0.1 dex.
In addition to providing new integral constraints on the cosmic star-formation rate density of galaxies in the EoR \citep[][see Fig.~\ref{fig:filter_example}]{Madau2014}, the JELS $z\sim6.1$ \halpha sample will make it possible to study the nature of star-forming galaxies at this epoch in a relatively unbiased sample and to constrain the scaling relations linking ongoing star formation to key galaxy properties such as masses, sizes (Stephenson et al. \emph{in prep}) and clustering properties (Hale et al., \emph{in prep}).

% To illustrate the complementarity between the JELS narrow-band and other slitless spectroscopy approaches, in Fig.~\ref{fig:halpha_sample_dist} we show the distribution in redshift and emission-line flux for a sample of the 35 most robust F466N/F470N \halpha emitters \citep[presented in][]{Pirie2024}), alongside the published sample of \halpha emitters detected in the FRESCO survey \citep{covelopaz2024}.
% \begin{figure}
%     \centering
%     \includegraphics[width=1\columnwidth]{figures/jels_fresco_z_fha_dist.pdf}
%     \caption{Illustration of the novel sample size and emission-line flux parameter space probed by the JELS \halpha selection compared to slitless spectroscopy samples in the literature. Filled hexagons show the photo-$z$ and measured line flux for the robust \halpha samples \citep[$F_{\text{line}}/ \text{erg\,s}^{-1} \text{cm}^{-2}$;][]{Pirie2024} selected in F466N and F470N. The FRESCO \halpha sample \citep{covelopaz2024} is shown for comparison. Light shaded regions correspond to the redshift range where \halpha falls within the minimum and maximum ($>1$ per cent of peak filter throughput) wavelengths covered by F466N/F470N, with the darker shaded ranges illustrating the corresponding effective wavelength range.}
%     \label{fig:halpha_sample_dist}
% \end{figure}
Given the high completeness within the survey volume and the range of line fluxes probed, the JELS \halpha sample also offers a valuable test-bed for understanding the potential biases and limitations of UV-based selection techniques that are ubiquitously used at high redshifts.
For example, the bursty nature of star-formation in low mass galaxies is now understood to play a key role in dictating the observability of high-redshift galaxies \citep{Sun2023a}, and hence significantly impacting on the inferred UV luminosity functions \citep{Sun2023b}, as well as strongly impacting estimates of key galaxy properties such as stellar mass \citep{Endsley2023} and ionising photon escape \citep{flury2025}.
With JELS, it is possible to derive the distribution of \halpha ($\sim10$ Myr time-scales) to UV star-formation rates ($\sim100$ Myr timescales) in a homogeneously selected sample of galaxies in the EoR for the first time.
By selecting on emission-line strength, the narrow-band selection is naturally sensitive to the highest equivalent width population and hence provides constraints on the youngest or most extreme burst populations \citep[see e.g.][]{Maseda2023}.
Recent studies with \emph{JWST} medium bands confirm the expectation that high-redshift galaxies displaying evidence for elevated recent star-formation activity are typically the most efficient ionising photon producers \citep[i.e. higher $\xi_{\text{ion}}$;][]{simmonds2024}, while modelling predicts that very high \halpha SFRs could correlate with increased ISM porosity \citep{clarke2002} and hence Lyman continuum photon escape.
The high EW emission-line sources selected by JELS may therefore offer an especially valuable probe of the most extreme ionisers of the IGM.

Further enhancing the scientific potential of the JELS \halpha sample in this area is the ongoing deep Multi Unit Spectroscopic Explorer \cite[MUSE;][]{Bacon2010} integral field spectroscopy over the full JELS survey footprint (ESO Large Programme 112.25WM.001, PI: Swinbank), reaching triple the exposure time per pointing of the MUSE-Wide Survey \citep{Urrutia2019} over $\sim1.75\times$ greater area.
The combination of a homogeneously selected \halpha sample with complete resolved rest-UV spectroscopy will enable a broad range of studies into the late stages of cosmic reionization, with the potential to improve on and complement existing studies of \Lya\, emitter fractions \citep{Stark2010}, \Lya\, emission-line profiles \citep{Mason2018}, constraints on $\xi_{\text{ion}}$ as a function of \Lya\, properties \citep{Prieto-Lyon2023,Saxena2024} and resolved studies of \Lya\, \citep[e.g.][]{Smith2018,Roy2023}.

 Another key advantage offered by narrow-band samples is that they are ideally suited for clustering analyses, as the narrow redshift slice minimises redshift projection effects \citep[cf.][]{Geach2012,Cochrane2017}.
Measurement of the correlation of \halpha emitters at $z\sim6.1$ will enable constraints on the dark-matter halos hosting these star-forming galaxies.
In addition to placing these sources into their broader cosmological context for direct comparisons with galaxy formation simulations, these measurements will offer another critical comparison against existing Lyman break and Ly$\alpha$ emitter galaxy samples at this epoch (see also Section~\ref{sec:goals-eor}). 

\subsubsection{The faint SF population at cosmic noon}\label{sec:goals-sfr}
One of the largest uncertainties in constraining the global cosmic SFR density is the precision to which the faint-end slope of respective LFs can be reliably constrained. 
The faint end of the UV LF has been tightly constrained across the bulk of cosmic history \citep[reaching $M_{\text{UV}} > -15$ out to $z > 6$;][]{Bhatawdekar2019, Bouwens2022, Harikane2022}, with very tight constraints at cosmic noon \citep[][$\alpha\pm0.04$]{Parsa2015} and clear evidence of steepening slopes as redshift increases.
Current constraints on the faint end of the \halpha LF at cosmic noon are, however, significantly more limited \citep{Sobral2012, terao2022}.
This is a critical measurement, since the difference between a faint-end slope of $\alpha = -1.75$ and $\alpha = -1.5$ corresponds to a factor-of-two difference in the integrated star-formation rate density.

The extreme depth of our \emph{JWST} NIRCam imaging means that the JELS F212N observations will detect \halpha emitters $\sim5\times$ fainter than previous ground-based studies, with an estimated sample size of $\sim200$.
The resulting determination of the faint-end slope of the $z=2.23$ \halpha luminosity function can achieve a precision of $\delta_{\alpha} < 0.05$, almost $3\times$ better than current limits, thereby tightly constraining the relative evolution of dwarf galaxies.
Furthermore, the extensive broadband imaging from HST/CANDELS and PRIMER will enable robust constraints on the stellar masses, star-formation histories and the dust attenuation of the \halpha sample. 

In addition to the F212N \halpha sample constraints at $z=2.23$, the JELS F466N/F470N filters also probe the Paschen lines \Pa and \Pb\, at $z\sim1.5$ and $2.6$ respectively, bracketing the peak of cosmic star-formation (Fig.~\ref{fig:filter_example}).
Essentially unaffected by dust, the \Pa\, (1.87$\mu$m) and \Pb (1.28$\mu$m)-lines offer unbiased instantaneous SFR-indicators \citep{cleri2022}.  
For the JELS F466N limiting line flux (90th percentile), the corresponding \Pa and \Pb luminosity limits reach $10^{40.25}$ and $10^{40.84} \text{erg s}^{-1}$ respectively ($0.17\,\text{dex}$ deeper in the central region).
Assuming Case B recombination, a temperature of 10,000K and electron density, $N_{e} = 10^{4}\,\text{cm}^{-3}$, we expect line intensity ratios of \halpha/\Pa $= 8.584$ and \halpha/\Pb $= 17.614$ \citep{Storey1995}.
The Paschen line sensitivities at $z\sim1.5$ (\Pa) and 2.6 (\Pb) therefore correspond to equivalent \halpha luminosities of $10^{41.18}$ and $10^{42.1} \text{erg s}^{-1}$, or star-formation rates of 0.35 and $3.6 \Msunpyr$ respectively\footnote{Where $\log_{10}(\mathrm{SFR}_{\text{H}\alpha}/\Msunpyr) = -41.64 + \log_{10}(L_{\text{H}\alpha}/ \text{erg\,s}^{-1})$ assuming the SFR calibration of \citet{Theios2019}.}.
The Paschen line depths are therefore comparable to, or deeper, than those achieved for existing \halpha narrow-band samples at these redshifts \citep{Geach2008, Sobral2013}.
We caveat that recent evidence suggests that the standard Case B assumption may not be valid in all galaxies, both in the lower redshift Universe \citep{Flury2022b, scarlata2024} and at the redshifts probed by the JELS Paschen samples \citep{Pirzkal2024}.
The inferred limiting star-formation rates are therefore only illustrative.
Statistical samples of bright Paschen line emitters at cosmic noon selected by JELS therefore also offer an ideal test-bed for future spectroscopic studies exploring the diversity of interstellar medium (ISM) conditions in star-forming galaxies through the distribution of Balmer and Paschen line ratios.
% Case B, 10000K, N=1e4: Ha/Pa = 8.584, Ha/Pb = 17.614
\begin{figure*}
    \centering
    \includegraphics[trim={0.2cm 0 0 0.2}, width=1\columnwidth]{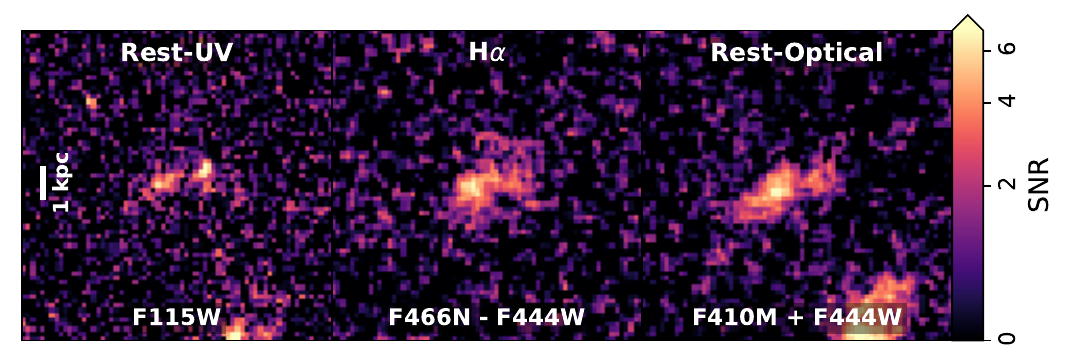}\vspace{-0.0cm}
    \includegraphics[trim={0.2cm 0 0 0.2},width=1\columnwidth]{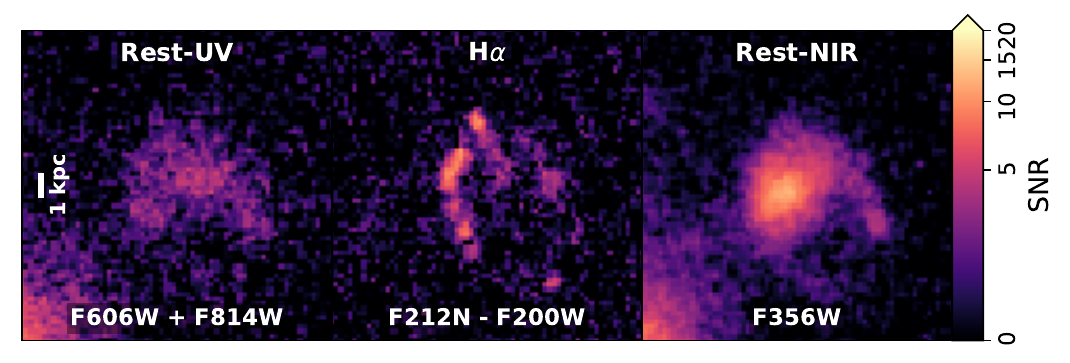}\vspace{-0.3cm}
    \includegraphics[trim={0.2cm 0 0 0.2},width=1\columnwidth]{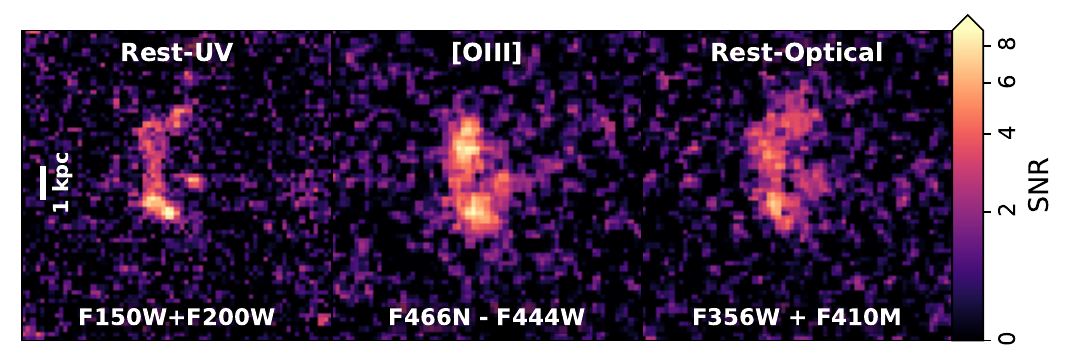}\vspace{0cm}
    \includegraphics[trim={0.2cm 0 0 0.2},width=1\columnwidth]{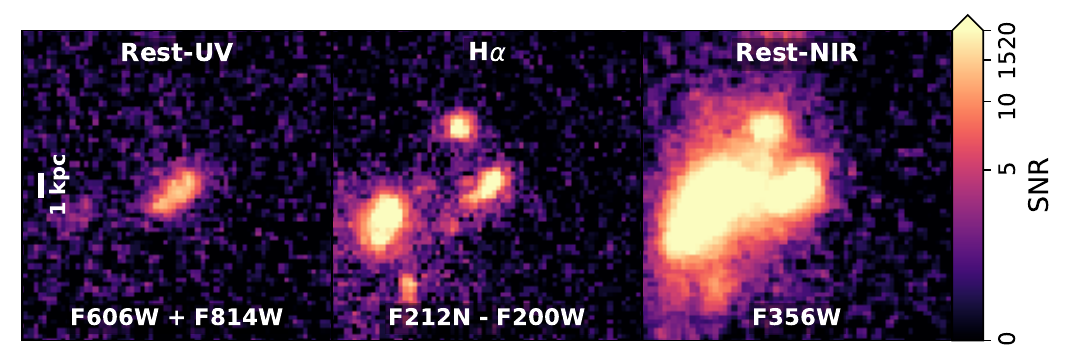}\vspace{-0.3cm}
    \includegraphics[trim={0.2cm 0 0 0.2},width=1\columnwidth]{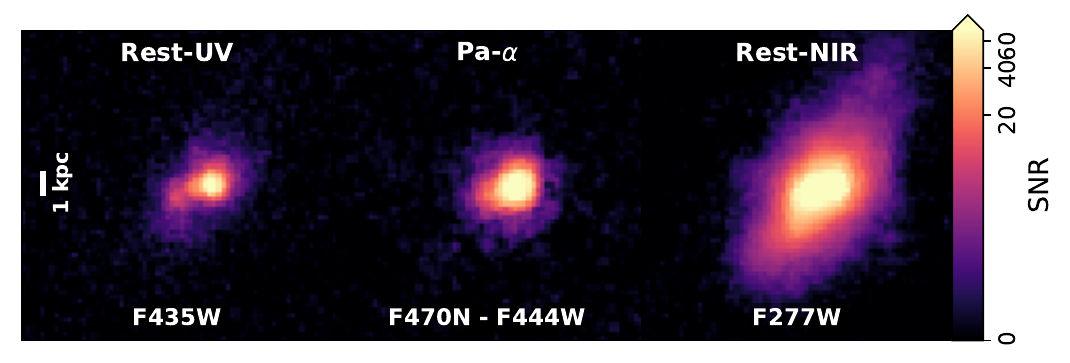}\vspace{-0.0cm}
    \includegraphics[trim={0.2cm 0 0 0.2},width=1\columnwidth]{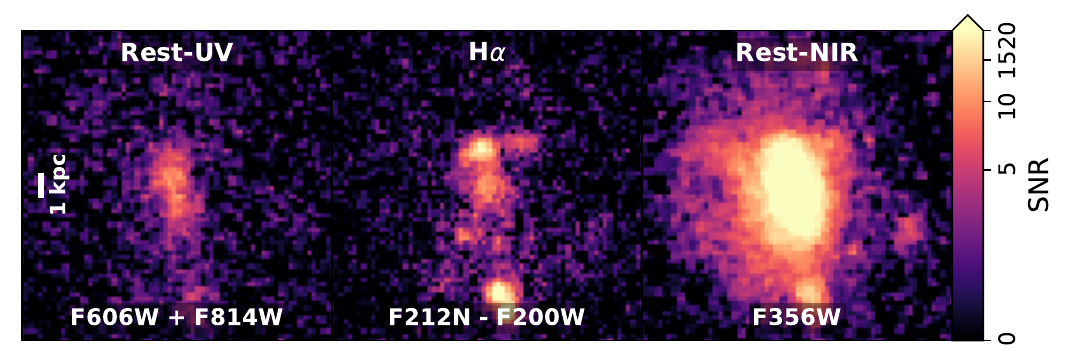}
    \caption{Illustration of resolved rest-UV (left), emission-line (centre) and rest-frame optical/NIR (right) for \halpha, \oiiia and \Pa emission line objects in JELS. Examples of F466N or F470N selected sources are shown in the left column, with examples of bright extended F212N excess sources shown in the right column. The corresponding broad and narrow-band filter combinations are indicated in each image. Cutouts are $2\,\textrm{arcsec} \times 2 \,\textrm{arcsec}$ in size, with consistent colour scales illustrated by the corresponding colour bar and 1 proper kiloparsec at the corresponding redshift shown for reference. }
    \label{fig:resolved}
\end{figure*}
\subsubsection{The role of galaxies in reionization -- \oiiia emitters at $8.3 \lesssim z \lesssim 8.5$}\label{sec:goals-eor}
Prior to JWST, the prevalence of high EW \oiiia emitters at $z > 6$ had been studied in large samples based on strong broadband colour excess \citep[e.g.][]{DeBarros2019}, with the \oiiia EW found to strongly correlate with redshift \citep{khostovan2016}.
Studies of the properties of \oiiia emitters at $z > 6$ from medium/broadband \emph{JWST} observations \citep{Simmonds2023, Begley2024, wold2025} and slitless spectroscopy \citep{Matthee2023A} indicate that the highest EW sources are likely to be producing ionizing photons with very high efficiency, and could therefore represent key drivers of the early stages of cosmic reionization.
Similar to the \halpha selection, homogeneously selected samples of \oiiia emission-line sources from JELS can therefore offer a probe of the earlier stages of reionization at $z\sim8.4$, both through the overall statistical constraints and by sign-posting the sites of the largest ionised bubbles that can then be probed through \Lya\ transmission modelling \citep[e.g.][]{Tang2024, Witstok2024}.
When the \oiii line falls within F466N ($z=8.298$ at $\lambda_{\text{pivot}}$), the narrow-band line flux limits presented in Table~\ref{tab:depths} correspond to luminosity limits of $L_{\oiii} \sim 10^{41.8}-10^{41.9}\,\text{erg s}^{-1}$ for the deeper JELS region, increasing to $\sim10^{42.0} \text{erg s}^{-1}$ for 90 per cent of the survey area.
At slightly higher redshifts, the fainter \oiiil line is also then redshifted into the narrow-bands (extending up to e.g. $z=8.492$ at $\lambda_{\text{pivot,F470N}}$) with comparable measured line sensitivity.
For an assumed intrinsic \oiii/\oiiil ratio of 2.98 \citep{Storey2000}, we note however the effective \oiiil sensitivity is $\sim0.5\,\text{dex}$ lower than for \oiii selection.
The JELS narrow-band imaging is highly complementary to slitless spectroscopic surveys, extending emission-line selections to fainter $L_{\oiii}$ at $z>8$ \citep[cf.][]{Meyer2024}, potentially improving constraints on the faint end of the \oiii LF and revealing key populations critical in the process of cosmic reionization.
Below we illustrate the exquisite morphological information provided by the narrow-band imaging (Section~\ref{sec:morph}) and the reliability and sensitivity of the selection (Section~\ref{sec:sims}).

\subsubsection{Other line emitter samples}\label{sec:goals-other}
Finally, while the science cases outlined above focus on intrinsically brighter emission-line species, the combination of high-sensitivity and extensive ancillary observations sufficient to distinguish between potential redshift solutions means that JELS is sensitive to a broad range of novel emission-line sources.
Further examples range from the 3.3\micron\, poly-aromatic hydrocarbon (PAH) feature at $z\sim0.4$ (F466N/F470N) in the lower redshift Universe, to \siii at $z\sim4.1$ (F466N/F470N) and \oii at $z\sim4.7$ (F212N) in the early Universe.
Sources with line strengths in intrinsically weaker optical--near-IR line species sufficiently bright to be robustly selected as excess sources in JELS likely represent ideal targets for detailed spectroscopic follow-up.
Regardless, the ability to isolate emission lines in narrow-band filters results in improved photometric redshift (photo-$z$) and SED modelling precision for all such sources, as well as the potential for detailed morphological studies.

\subsection{Spatially resolving ionised gas properties}\label{sec:morph}
Early \emph{JWST} observations have demonstrated that the preceding picture where regular Hubble-sequence morphologies emerged around cosmic noon \citep[$1 < z < 3$;][]{Mortlock2013} may not be correct, with discy morphologies potentially dominating the galaxy population as early as $z \sim 7$ \citep[e.g.][]{Ferreira2022, Kartaltepe2023, Conselice2024}.
Robustly measuring galaxy morphologies at $z > 2$ is crucial for far more than simple galaxy classification.
Constraining the spatial distribution of on-going star-formation within galaxies as a function of stellar mass (and ideally halo mass), or other key properties such as AGN activity, can directly inform models of feedback in hydro-dynamical simulations \citep{Cochrane2023}.
One of the key advantages offered by narrow-band emission-line selection over slitless spectroscopic surveys is immediate access to the robust 2D emission-line morphologies in \emph{individual} galaxies.
This means that JELS will enable studies of resolved ionised gas properties in less biased galaxy samples on sub-kpc scales without the need for complex forwarding modelling of multiple dispersion directions \citep{Pirzkal2018, Estrada-Carpenter2024, Shen2024}, or the stacking analysis of statistical samples \citep{Nelson2013,Matharu2024,Liu2024}.

As with integrated measurements above (Sections~\ref{sec:goals-halpha}-\ref{sec:goals-other}), this is particularly valuable when the ionised gas offers a direct SF tracer. The JELS \halpha and F466N/F470N \Pa/\Pb samples offer a clean probe of sub-L$^\star$ galaxies at their respective redshifts whose multi-wavelength properties, resolved structures and parametrised morphologies can be compared with those of brighter galaxies (and lower-redshift samples) to investigate the physical processes driving star formation within these galaxies \citep[e.g.][]{Cochrane2021}. 
In Fig.~\ref{fig:resolved} we show the rest-UV, ionised gas (narrow-band excess) and rest-optical/near-infrared continuum morphologies for examples of both F466N/F470N and F212N excess selected emission-line galaxies.
The advantage of resolving both the UV and \halpha (or \Pa) star formation is immediately evident, with significant variation both between the two star formation probes and the underlying continuum that gives insights into the star-formation properties of galaxies that cannot be obtained from one alone (e.g. the distribution of dust, or the variation of star-formation timescales within the galaxy).
When extended to the full narrow-band samples, JELS can therefore constrain the morphology of on-going star formation compared to that of the in-situ stellar mass (measured from resolved SED fitting with full PRIMER observations) for a representative sample of SFGs, testing whether the inside-out growth of galaxies inferred from stacks at $z < 1.5$ \citep{nelson2016} is true for all individual galaxies and tracing this over a critical period in the morphological history of galaxies. 

\section{Predicted \halpha and \oiii redshift and emission line recovery}\label{sec:sims}
Demonstrating the relevant sensitivity and physical constraints enabled by the JELS narrow-band imaging for all of the potential science cases is impractical given the wide range in emission lines, redshifts and associated physical properties (integrated or resolved) probed.
Nevertheless, for the key target samples of \halpha and \oiii-emitters, it is informative to test the practical ability for JELS to recover emission line selected galaxies in narrow redshift slices at $z > 6$, alongside the precision to which corresponding emission line properties can be estimated.
Additionally, we can explore the quantitative advantages offered by narrow-band imaging over existing broadband imaging alone, or potential alternatives such as medium-band surveys that could probe larger cosmological volumes at the expense of redshift precision. 

To generate a range of intrinsic SEDs that span a plausible range in equivalent widths while also providing realistic accompanying continuum and broadband colours, we use the \texttt{Prospector} Bayesian SED modelling code \citep{Prospector} to efficiently generate stellar population models from the Flexible Stellar Population Synthesis \citep[FSPS;][]{Conroy2009, Conroy2010,pythonFSPS} package, with accompanying \texttt{Cloudy} \citep{Cloudy} photo-ionization nebular line and continuum emission as outlined in \citet{Byler2017}.

To avoid unnecessary duplication and analysis, we make the simplifying assumption that any conclusions drawn from an F466N sample will be largely applicable to an F470N-selected sample, or that the increased sensitivity of F470N would only result in increased precision or sensitivity.
For each of the emission line samples, we therefore generate a full parent sample of mock galaxies over a range of redshifts corresponding to just the JELS F466N wavelength coverage. 
At each redshift step, we generate 1000 mock galaxy SEDs with a range of stellar population parameters designed to ensure that the emission line equivalent widths extend below the range expected for real galaxies samples at these redshifts \citep[e.g.][]{Endsley2023b}.
The redshift ranges and corresponding ages for each sample are chosen to be:
\begin{itemize}
    \item \halpha: Redshifts spanning $6.06 < z < 6.12$ with a step size of $\delta z=0.01$. For all redshift steps, the time since the onset of star-formation, $t_{\text{age}}$, for a constant star-formation history (CSFH) is drawn from a log-uniform prior in the range $-2.3 < \log_{10}(t/\text{Gyr}) < -0.15$. 
    \item \oiii: Redshifts in the range of $8.25 < z < 8.35$ with $\delta z=0.01$, and a corresponding CSFH $t_{\text{age}}$ distribution from $-2.3 < \log_{10}(t/\text{Gyr}) < -0.37$.
\end{itemize}

For both mock galaxy samples, metallicity is drawn from a flat distribution in $\log_{10}(Z/Z_{\odot})$ in the range of $-0.7 < \log_{10}(Z/Z_{\odot}) < -0$ (with the assumption that $Z_{\text{gas}} = Z_{\text{stellar}}$).
Additionally, to allow for a plausible variation in emission line properties the ionisation parameter, $\log_{10}(U)$, is drawn from a range appropriate for the metallicity \citep[$-3 < \log_{10}(Z/Z_{\odot}) < -2$, see e.g.][]{reddy2023}.
For dust attenuation, we assume a simple \citet{Calzetti2000} dust attenuation law with extinction magnitude, $A_{V}$, drawn from a log-normal distribution with a mean of $\ln(A_{V}) = -1.69$ (i.e. $A_{V} = 0.18$) and standard deviation of 0.5.
The specific distribution values are chosen to broadly to match the typical attenuation estimated for the real observed \halpha emitter sample \citep{Pirie2024}.

For the simulated SED population, we generate the corresponding intrinsic flux densities in the HST Optical (ACS/WFC F435W, F606W and F814W), PRIMER NIRCam and JELS F466N/F470N filters \citep[see Table 1 of][]{Pirie2024}.
Additionally, to enable simulation of an equivalent medium-band emission line selection survey, we also generate photometry in the NIRCam/F460M ($\lambda_{\text{pivot}}=4.63\micron\,$, $W_{\text{eff}}=0.23\micron$) that encompasses the JELS LW narrow-band filters.
The `true' \halpha/\oiii emission line luminosities and equivalent widths are measured directly from the corresponding noise-free model spectra. 
We note that the \halpha measurements do not account for underlying Balmer absorption, however for the purposes of this experiment we determine that any resulting systematics are negligible on the basis that the young stellar populations and corresponding high equivalent widths probed by these high redshift sources means such corrections will typically be minimal.
With the emission line properties measured, the parent SED sample and the corresponding mock photometry is renormalised to a constant line luminosity of $L_{\text{line}} = 10^{41}\text{erg\,s}^{-1}$.

For the subsequent analysis, the mock galaxy samples and associated photometry are then scaled to create 10 different samples with intrinsic emission line luminosity (\halpha, \oiii) spanning the ranges of $41 \leq \log_{10}(L_{\text{H}\alpha}/\text{erg\,s}^{-1}) \leq 43$ and $41.3 \leq \log_{10}(L_{\oiii}/\text{erg\,s}^{-1}) \leq 43.3$ respectively.
Photometric uncertainties for each filter are drawn from a Gaussian distribution with a standard deviation set by the median uncertainties measured for faint sources in 0.3\arcsec diameter apertures in the PSF-homogenised images and associated catalogues presented by \citet{Pirie2024}.
For simplicity, we assume that the mock galaxies are point sources and for the target PSF in the JELS PSF-homogenised photometry catalogues, the 0.3\arcsec aperture captures 50.3\% of the total flux of a point-source.
We therefore correct the simulated SNR accordingly.

For the simulated F460M filter, for which current imaging does not exist, we assume a typical depth that is twice as sensitive as measured in the existing PRIMER F410M imaging.
Given that the instrumental sensitivity for F410M is approximately twice that of F460M for the same equivalent exposure time\footnote{See e.g. \href{https://jwst-docs.stsci.edu/jwst-near-infrared-camera/nircam-performance/nircam-sensitivity}{https://jwst-docs.stsci.edu/jwst-near-infrared-camera/nircam-performance/nircam-sensitivity}}, the F460M sensitivity for our mock medium-band survey therefore corresponds to $\sim16\times$ the PRIMER F410M exposure time and equivalent to observing for the combined F466N and F470N exposure times.

% \begin{itemize}
%     \item Photo-$z$ method
%     \item Photo-$z$ results discussion:

%     \item Line flux method
%     \item Line flux results discussion:
% \end{itemize}

\subsection{Photometric redshift precision of narrow-band selected samples}\label{sec:sims-photoz}
To explore the practical photo-$z$ precision achieved for a JELS narrow-band selection, we first run the mock emission line galaxy sample through photo-$z$ analysis comparable to that applied to either real JELS samples or other JWST-selected broadband samples. 
To enable like-for-like comparison, we perform photo-$z$ analysis for the mock sample using three different subsets of JWST NIRCam filters:
the full PRIMER filter-set plus JELS LW narrow-band filters (F466N/F470N), the PRIMER filters supplemented with the deep F460M medium-band filter, and a PRIMER-only run. The same HST optical filters (F435W, F606W and F814W) are included for all three runs. 

We measure photo-$z$s using the EAzY \citep{brammer2008} template fitting code using the default \texttt{fsps} set supplemented with the high-$z$ appropriate templates of \citet{larson2023b}.
Since the photometry is completely synthetic, we do not perform any photometric zeropoint offset corrections as part of the photo-$z$ analysis.
However, we do include an additional 5\% systematic uncertainty (added in quadrature) during the template fitting analysis.
We quantify the precision of the resulting photo-$z$ estimates as a function of intrinsic line luminosity and rest-frame equivalent width using two metrics. 
Firstly, we define the metric, $\Delta_{z}$, as the redshift range that spans between the 16th to 84th percentiles of the stacked photo-$z$ posterior of all galaxies in a given bin, i.e. $z_{84} - z_{16}$ where for example $z_{84}$ is defined such that
$\int_{0}^{z_{84}} \text{CDF}(z)\,\text{d}z = 0.841$.
Secondly, we calculate the robust scatter of a given bin, $\sigma_{\textup{NMAD}}$, defined as:
\begin{equation}
\sigma_{\textup{NMAD}} =1.48 \times \text{median} ( \left | \delta z \right | / (1+z_{\textup{true}})),
\end{equation}
where $\delta z = z_{\textup{phot}} - z_{\textup{true}}$.
Together, these metrics give a picture of both the typical precision to which an individual galaxy's redshift can be constrained ($\Delta_{z}$) and the overall sample precision ($\sigma_{\textup{NMAD}}$).
Note that the \emph{catastrophic} outlier fraction (i.e. $\delta z / (1+z_{\textup{true}}) > 0.15$) was also investigated as a metric. 
However, for the majority of the emission line parameter space probed, in the narrow-band and medium-band photo-$z$ runs the outlier fractions were exactly zero such that any quantitative comparison was uninformative.

\begin{figure*}
    \centering
        \includegraphics[width=0.95\textwidth]{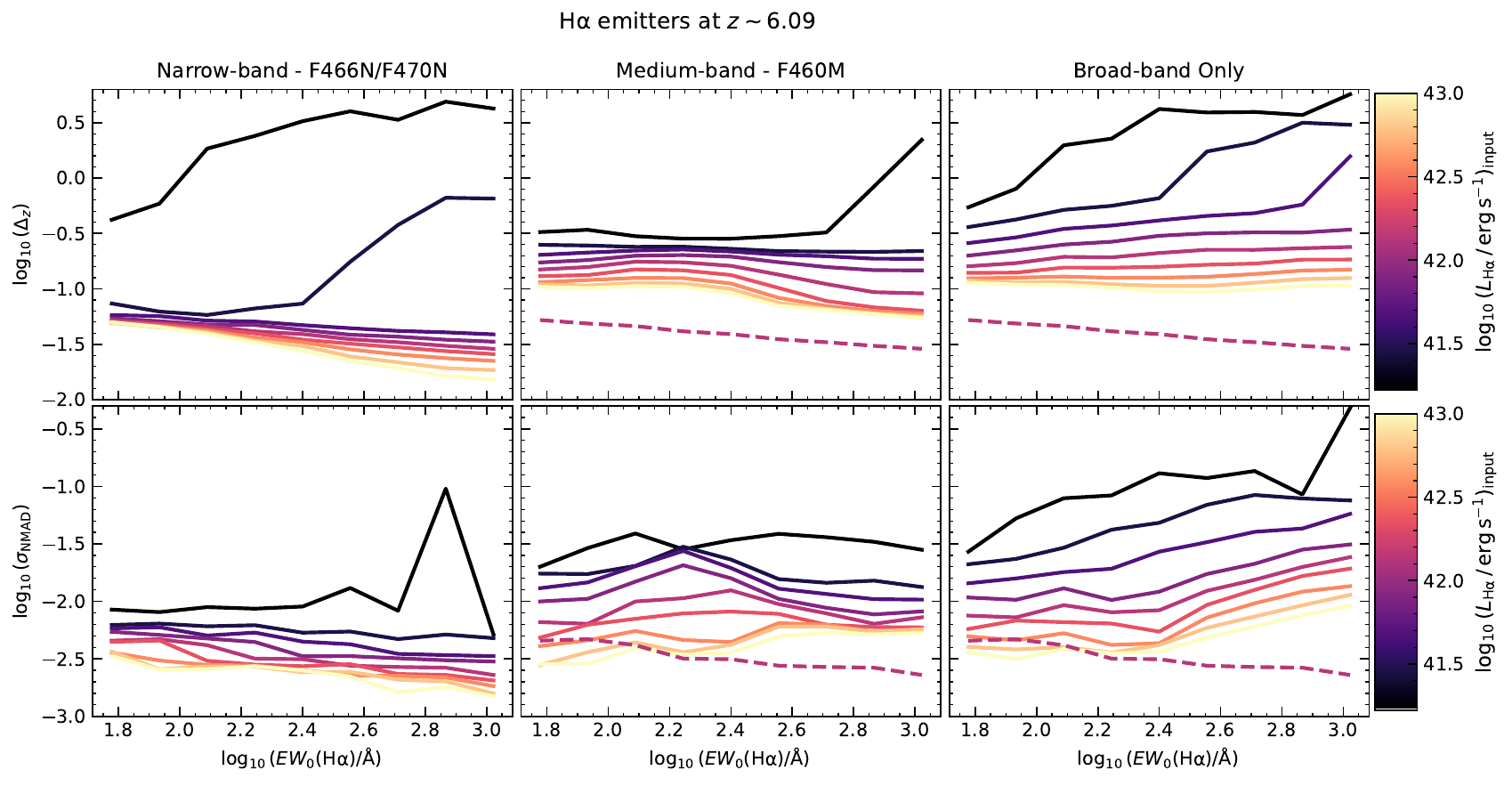}
        \includegraphics[width=0.95\textwidth]{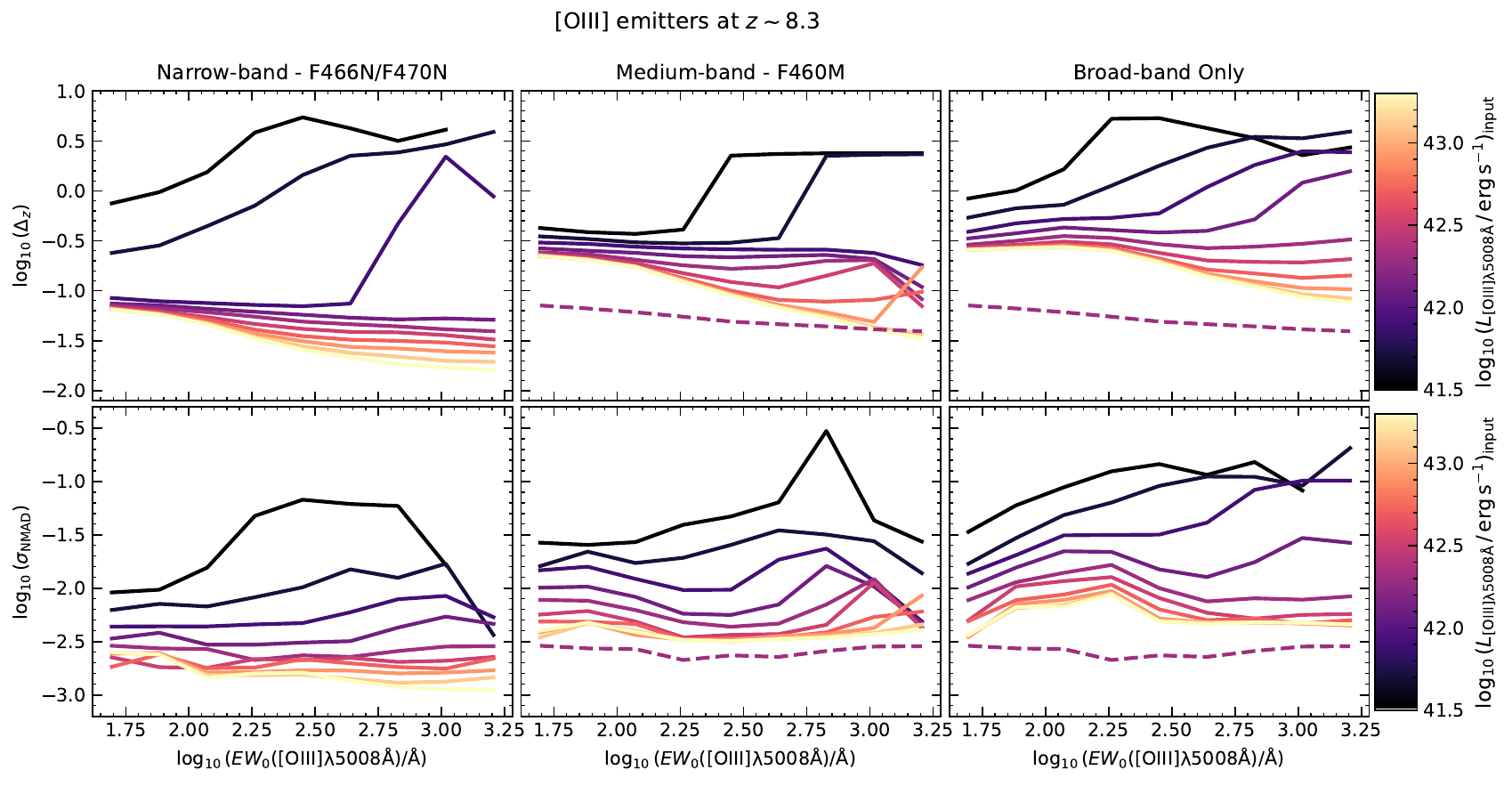}
    \caption{Photo-$z$ statistics, $\Delta_{z}$ and $\sigma_{\textup{NMAD}}$, as a function of rest-frame equivalent width $\text{EW}_{0}$ ($x$-axes) and intrinsic line-luminosity (colour-scale) for the simulated \halpha (top) and \oiii (bottom) emission-line galaxy samples. For each sample, we present the statistics when both JELS LW narrow-bands are included, when extremely deep F460M medium-band filter is included, and when using the PRIMER filters only (left, centre and right columns respectively). To aid comparison, on the middle and right-hand columns the dashed line illustrates the corresponding narrow-band sample statistics for an intrinsic line luminosity close to $L^{\star}$ (specifically the bins centred at $\log_{10}(L_{\text{H}\alpha}/\text{erg\,s}^{-1})\sim42$ and $\log_{10}(L_{\oiii}/\text{erg\,s}^{-1})\sim42.3$).}
    \label{fig:photoz_stats}
\end{figure*}

In Fig.~\ref{fig:photoz_stats} we present the resulting photo-$z$ statistics for both the simulated \halpha and \oiii emission-line samples. 
To determine whether a given source would be included in the corresponding photometric sample, we apply a SNR cut similar to that applied in real sample selections \citep{Pirie2024}.
For all three filter sets we require $f_{\text{F444W}}/\sigma_{\text{F444W}} > 5$ as well as $f_{\text{F356W}}/\sigma_{\text{F356W}} > 3$.
Additionally, for the narrow-band and medium-band runs we require $f_{\text{F466N}}/\sigma_{\text{F466N}} > 5$ and $f_{\text{F460M}}/\sigma_{\text{F460M}} > 5$ respectively.

We observe a number of trends in photo-$z$ precision as a function of intrinsic properties that are consistent across both emission-line samples.
Firstly, there is a consistent correlation between increased emission-line luminosity and improved photo-$z$ precision in both metrics (i.e. lower $\sigma_{\textup{NMAD}}$ and $\Delta_{z}$).
This correlation is to be expected as the line-luminosity most directly correlates with photometric signal-to-noise.
Secondly, both the individual and ensemble photo-$z$ precision achieved when JELS narrow-band filters are included is typically $\approx2-5\times$ better than the mock medium-band survey for the equivalent intrinsic properties (cf. the effective width of the F460M filter being $\approx4\times$ wider than the F466N filter).
The improvement in $\Delta_{z}$ gained from the medium-band filter when compared to PRIMER-only is less significant, while for a large fraction of parameter space the medium-band survey yields no significant improvement in $\sigma_{\textup{NMAD}}$ for the same comparison.

We do, however, see variations in the relative improvement when including a narrow-/medium-band, as a function of $\text{EW}_{0}$.
For low rest-frame equivalent widths ($\text{EW}_{0} < 200\text{\AA}$), the dependence of $\Delta_{z}$ on line luminosity becomes negligible.
By construction, since the input SEDs are normalised to a given line-luminosity, lower equivalent width sources will have correspondingly brighter stellar continuum and hence higher SNR in their broadband photometry.
The convergence in photo-$z$ statistics at lower-$\text{EW}_{0}$ when narrow-band filters are included illustrates the parameter space where the photo-$z$ precision is no longer dictated by the precision to which the relevant emission line can be constrained, but instead dominated by other broadband features such as the Lyman and Balmer breaks.
In this regime, there is still, however, a significant improvement from the inclusion of narrow-band, while corresponding medium-band estimates offer only marginal gains over broad-band only estimates.
The majority of $z > 6$ galaxies are expected to have significantly stronger emission-line contributions;
\citet{Endsley2023b} estimate the typical $\text{EW}_{0}(\text{H}\alpha)$ for $z\sim6$ galaxies ranges from $\sim580-850\text{\AA}$ ($\log_{10}(\text{EW}_{0}(\rm{H\alpha})/\text{\AA}) \sim 2.8$).
Fig.~\ref{fig:photoz_stats} therefore demonstrates that for the expected emission line luminosities probed by JELS \halpha and \oiii samples (see Table~\ref{tab:depths}), the inclusion of narrow-band filters should yield samples with typical photo-$z$ scatter $\sigma_{\textup{NMAD}} \lesssim 0.005\times(1+z)$ and with individual photo-$z$s constrained to $\Delta_{z} \lesssim 0.03$.
Furthermore, we note that for input line luminosities of $L_{\textsc{Oiii},\text{input}} > 10^{42}\,\text{erg s}^{-1}$, the JELS narrow-bands are able to constrain the redshift in individual sources to $\Delta_{z}<0.08$.
Compared to the redshifts of \oiiil and \oiii at the pivot wavelength of the F466N narrow-band filter ($z=8.29$ and $z=8.38$ respectively), this illustrates that the combination of both JELS narrow-bands and PRIMER photometry enables sufficient photo-$z$ precision to reliably identify and isolate individual \oiiia lines.
In contrast, the medium-band photo-$z$ estimates are only able to achieve such precision for the most luminous and highest-EW \oiii-emitters.

The increased precision from NB-based photo-$z$s is particularly advantageous for enabling efficient follow-up observations with ALMA to study far-infrared molecular lines \citep[e.g.][]{Carniani2017, LeFevre2020, Bouwens2022}.
The predicted photometric precision corresponds to $\pm$\,600–1700 km\,s$^{-1}$, sufficient to ensure the redshifted [C{\sc ii}]\,158\,$\mu$m or [O{\sc iii}]\,88\,$\mu$m lines would fall into a 3.75-GHz ALMA side band in band 6 or 7, respectively.
In Section~\ref{sec:examples}, we further illustrate the practical photo-$z$ precision achieved with real JELS selected emission-line galaxies for which spectroscopic redshifts have been obtained.

% \begin{figure*}
%     \centering
%         \includegraphics[width=0.95\textwidth]{figures/lhalpha_eqw_pzstats_allsets.pdf}
%         \includegraphics[width=0.95\textwidth]{figures/loiii_eqw_pzstats_allsets.pdf}
%     \caption{}
%     \label{fig:photoz_stats}
% \end{figure*}

\subsection{Sensitivity and accuracy of emission line recovery}\label{sec:sims-lums}
Building on the results above, it is also instructive to explore the corresponding accuracy of the emission-line luminosity that can be recovered from the JELS narrow-band excess emission. 
% For the simulated emission-line galaxy samples, we therefore apply narrow-band excess selection criteria closely following the approach applied by \citet{Pirie2024} for selecting \halpha emitters. 
For a simulated galaxy to be `selected' as an emission line excess source, we require the same individual filter SNR cuts as outlined above in Section~\ref{sec:sims-photoz}, an emission-line excess significance of $\Sigma > 3$ \citep[Eq.\,2 of][]{Pirie2024} and that over 50\% of the photo-$z$ posterior lies within the range $6.06 < z < 6.12$ for \halpha emitters or $8.25 < z <8.35$ for \oiii emitters. 
For \halpha emission line fluxes from the simulated photometry, we follow the standard prescription for estimating the line-flux based on the observed narrow-band excess \citep{Sobral2013, Pirie2024}:
\begin{equation}
\label{eq:f_line_1}
F_{\rm{H}\alpha} \ = \ \Delta \lambda_{\rm{F466N}} \ \frac{f_{\lambda,\rm{F466N}} \ - \ f_{\lambda,\rm{F444W}}}{1 \ - \ (\Delta \lambda_{\rm{F466N}} / \Delta \lambda_{\rm{F444W}})}\,\text{erg s}^{-1} \text{cm}^{-2},
\end{equation}
where $f_{\lambda,i}$ is the measured flux density in filter $i$, in units of $\text{erg s}^{-1} \text{cm}^{-2} \text{\AA}^{-1}$ and the corresponding filter effective widths, $\Delta \lambda_{i}$ in $\text{\AA}$.
We also subtract a constant $0.021\,\text{dex}$ correction to account for the contribution of \nii\, to the narrow-band emission, both for consistency within this manuscript (cf. Section~\ref{sec:lw_props}) but also to illustrate the validity of this approach in a more realistic application.
For \oiii, a similar approach can be taken. 
However, due to the very narrow rest-frame wavelength probed by the JELS narrow-bands at $z\sim8.3$, accurately estimating the \oiii luminosity requires correcting for an additional contribution from \oiiil that contributes to the narrow-band flux only at some redshifts.
While small, these corrections are non-negligible given the high emission line equivalent widths of galaxies at this redshift. 
In Appendix~\ref{sec:app-oiii_corr}, we outline how we account for these corrections and the changes required to Eq.~\ref{eq:f_line_1} for measuring \oiii line luminosities.

\begin{figure}
    \centering
        \includegraphics[width=0.99\columnwidth]{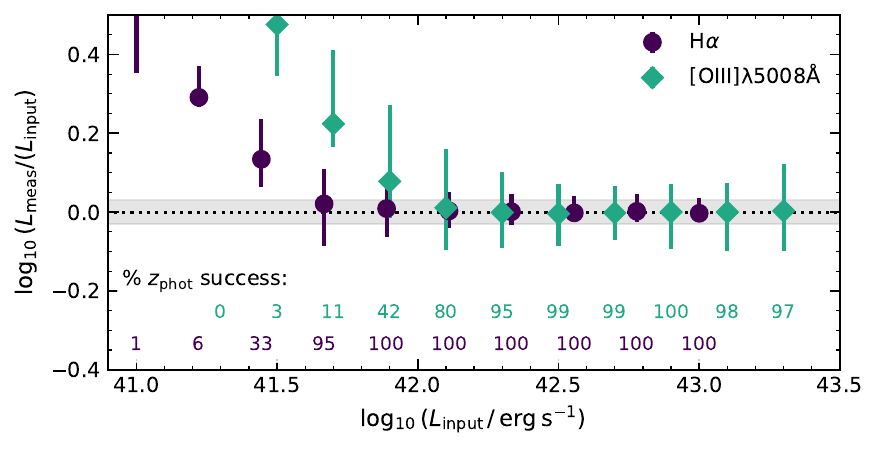}
    \caption{Measured emission line luminosities, $L_{\text{meas}}$, as inferred from the F466N narrow-band excess relative to the true intrinsic line luminosities, $L_{\text{input}}$, for the simulated \halpha (purple circles) and \oiii (teal diamonds) samples. For each bin in input luminosity, we show the median and 16 to 84th percentiles of measured to input line luminosity for a random sub-sample weighted to produce the expected EW distribution (see text). The shaded horizontal region illustrates $0\pm0.02\text{dex}$. Also shown are the fraction of simulated galaxies in each weighted sub-sample that pass the photo-$z$ selection criteria. Altogether, this figure illustrates that for emission lines above the 5$\sigma$ limits (see Table~\ref{tab:depths}), JELS can reliably select $>95\%$ of line-emitters and accurately measure the apparent line luminosity with negligible bias and with typical precision of 5 (\halpha) to 10\% (\oiii).}
    \label{fig:eline_recovery}
\end{figure}

As with the photo-$z$ simulation above, our mock emission line galaxy sample allows us to calculate the accuracy of narrow-band estimated emission line luminosities as a function of both true input luminosity and rest-frame equivalent width.
For the purpose of visualisation, and to provide an illustration of the realistic average accuracy, we instead choose to estimate the average accuracy marginalised over the expected $\text{EW}_{0}$ distributions for \halpha at $z\sim6$ and \oiii at $z\sim8$.
We base our assumed $\text{EW}_{0}$ distributions on the observational results from JWST photometric samples, specifically using the inferred \halpha $\text{EW}_{0}$ distribution for `faint' galaxies from \citet{Endsley2023b} with a log-normal distribution of mean $\mu_{\text{EW}_{0}} = 580\text{\AA}$ and width $\sigma_{\text{EW}_{0}} = 0.26\,\text{dex}$.
For \oiii, we assume a log-normal distribution with $\mu_{\text{EW}_{0}} = 380\text{\AA}$ and $\sigma_{\text{EW}_{0}} = 0.4\,\text{dex}$, broadly consistent with the \oiiia+\hbeta\, $\text{EW}_{0}$ distributions inferred by \citet{Endsley2023b} and \citet{Begley2024} with the mean scaled based on the assumed ratio of \oiii/\oiiil $=2.98$.
For each input luminosity bin we make 100 draws from the photo-$z$ posterior of each source (see Appendix~\ref{sec:app-oiii_corr}) in the bin before randomly selecting 500 samples with a probability weighting based on the assumed $\text{EW}_{0}$ distribution.

In Fig.~\ref{fig:eline_recovery} we present the resulting distribution of measured to `true' line luminosity measured directly from the original noise-free simulated spectrum.
For each luminosity bin, we show the median measured to input luminosity ratio, with error bars indicating the 16 to 84th percentiles of the distribution.
We see that for both emission line samples, at bright intrinsic luminosities the narrow-band estimated luminosities are measured to both a high accuracy (with bias $<0.01$ dex) and to good precision, with a scatter of $\sim0.02-0.05$ dex for \halpha and $\sim0.05-0.1$ dex for \oiii.
For input luminosities close to and below the expected 5$\sigma$ detection limit (see e.g. Table~\ref{tab:depths}), the average measured luminosity becomes significantly biased relative to the input population. 
The bias arises from the fact that the intrinsic fluxes for the faint sources are near or below the flux limits for the narrow-band filter, so the subset of sources that pass the selection criteria are those that are scattered above the detection threshold and hence the resulting average inferred luminosity is biased high.
To support this conclusion, in Fig.~\ref{fig:eline_recovery} we also show the percentage of the simulated sample (again weighted by $\text{EW}_{0}$) in each luminosity bin that passes the individual SNR, excess and photo-$z$ selection criteria. 
For the luminosity bins exhibiting significant bias ($L_{\text{H}\alpha,\text{input}} < 10^{41.5}\,\text{erg s}^{-1}$, $L_{\textsc{Oiii},\text{input}} < 10^{42}\,\text{erg s}^{-1}$), the fraction of galaxies passing the selection criteria is below 50 per cent.

\subsection{Summary of simulation results}
Together, the simulation results for photo-$z$ precision and line recovery presented here illustrate that for any emission-line sample with luminosities above the 5$\sigma$ detection limit, the photo-$z$s should be constrained to extremely high precision ($\sigma_{\text{NMAD}} <0.005\times(1+z)$). 
The inferred narrow-band luminosities measured from JELS will also be both accurate ($\pm0.02-0.1\,\text{dex}$) and precise ($\pm0.01\,\text{dex}$), with the uncertainties for scientific analysis therefore likely dominated by factors such as the dust attenuation corrections required to estimate intrinsic emission-line luminosities -- a limitation true also true for slitless spectroscopic samples and many spectroscopic surveys.

We caveat however that these simplified simulations do not account for realistic extended morphologies, which will decrease the precision of any estimates through both increased noise and systematic uncertainties from measuring \emph{total} fluxes.
Future JELS studies will incorporate more extended completeness simulations that account for the full range of observed morphologies and sizes (Pirie et al., \emph{in prep}).
Furthermore, we have presented simulation results only for sources where the line species of interest falls within the target JELS narrow-band.
Therefore, while we have demonstrated that the JELS narrow-band filters provide substantial gains in photo-$z$ accuracy within the target redshifts, JWST medium-band surveys will naturally offer the advantage of probing the high-$\text{EW}_{0}$ lines over a wider redshift range \citep[e.g.][see also \citeauthor{minerva_prop}~\citeyear{minerva_prop}]{suess2024} and provide complementary constraints to JELS.

\section{Spectroscopically confirmed JELS emitters}\label{sec:examples}
To further illustrate the diversity of galaxy properties present within the populations selected by JELS and the efficacy of the narrow-band selection, here we present spectroscopic confirmations of four high-$z$ line emitters selected by JELS that were included as filler targets in the Director's Discretionary program DD 6585 (PI: Coulter).

\begin{figure}
    \centering
    \includegraphics[width=0.98\columnwidth]{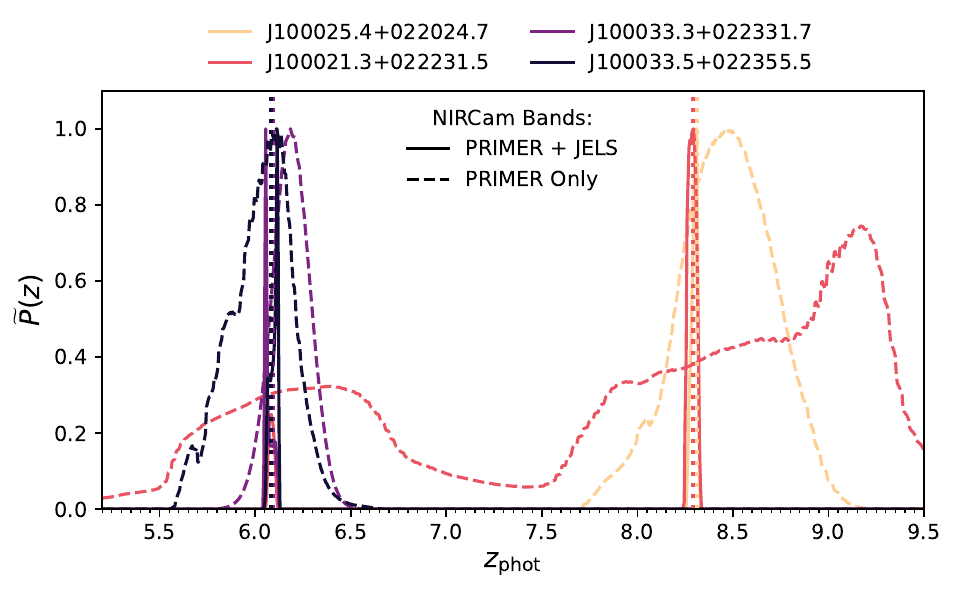}
    \caption{Normalised photo-$z$ posteriors, $\widetilde{P}(z)$, with (solid lines) and without (dashed lines) the inclusion of the JELS F466N and F470N filters for the spectroscopically confirmed line emitters. Due to the extremely narrow posteriors when JELS narrow-band filters are included, the photo-$z$ posteriors have been normalised by their maximum a posteriori values to aid visual comparison. Spectroscopic redshifts from the DDT Prism observations are illustrated by corresponding vertical dotted lines.}
    \label{fig:pofz_nb_bb}
\end{figure}

The JELS sample configured in the NIRSpec PRISM observations was selected from a F466N detected catalogue with $\text{SNR}_{\text{F466N}} > 5$ in 0.3 arcsec apertures. 
All sources satisfy emission-line excess criteria with colours $F444W-F466N$ > 0.3 and $F470N-F466N$ > 0.15 (corresponding to a rest-frame equivalent width of $\text{EW}_{0} \gtrsim 20 \text{\AA}$) and with emission-line excess significance, $\Sigma > 3$ \citep[see Eq.~2 of][]{Pirie2024}.
Photo-$z$s derived from PSF homogenised photometry from all available \emph{HST}/ACS, \emph{HST}/WFC3 and \emph{JWST}/NIRCam filters identified all sources as secure high-$z$ ($z > 5.5$) candidates\footnote{The photo-$z$s were estimated using EAzY \citep{brammer2008} with three different template sets: the default \texttt{fsps} set supplemented with the high-$z$ templates of \citet{larson2023b}, the \texttt{sfhz} set supplemented with the obscured AGN template of \citet{Killi2023} and the \texttt{eazy\_v1.3} set. The consensus photo-$z$s combining all three estimates are derived following the procedure outlined in \citet{Duncan2019}. See \citet{Pirie2024} for full details.}.
The resulting photo-$z$ posteriors are shown in Fig.~\ref{fig:pofz_nb_bb} (solid lines), along with the corresponding photo-$z$ posteriors for the same sources when only the PRIMER NIRCam photometry is used (dashed lines).
Of the four JELS emission-line candidates observed with NIRSpec, two are robustly identified as \halpha emitters with well constrained photo-$z$s at $z\sim6.1$, with the other two robustly identified as $z\sim8.3$ \oiii emitters.

\begin{figure*}
    \centering
        \includegraphics[width=0.98\columnwidth]{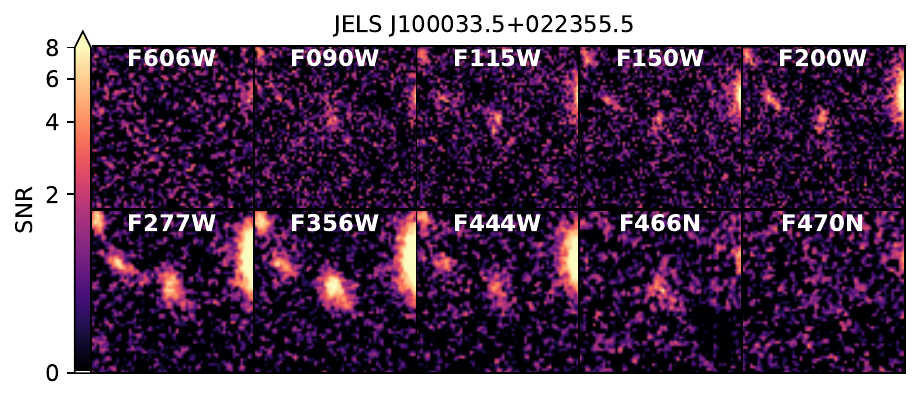}
    \includegraphics[width=0.98\columnwidth]{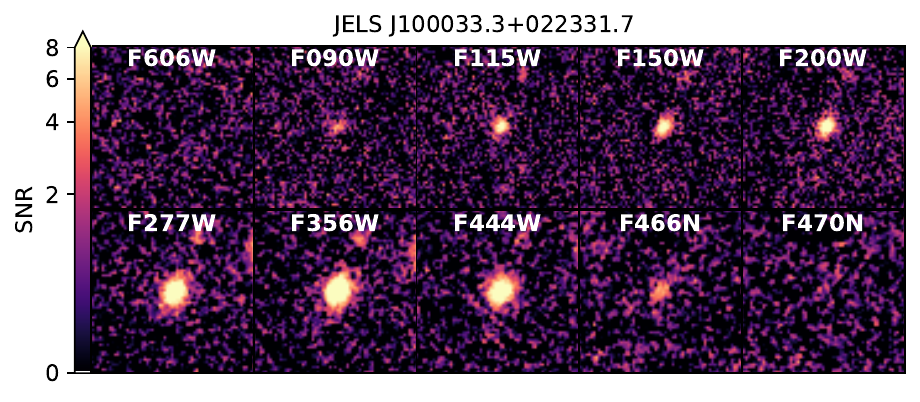}\vspace{-0.3cm}
    \includegraphics[width=0.98\columnwidth]{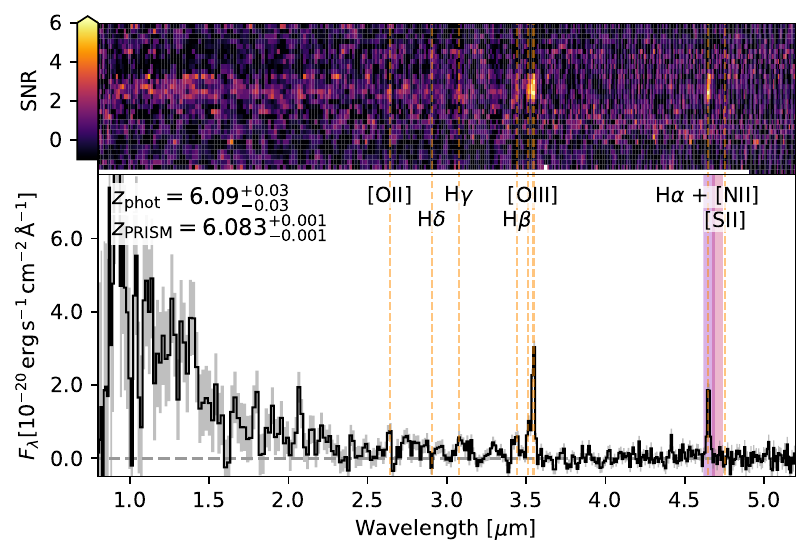}
\includegraphics[width=0.98\columnwidth]{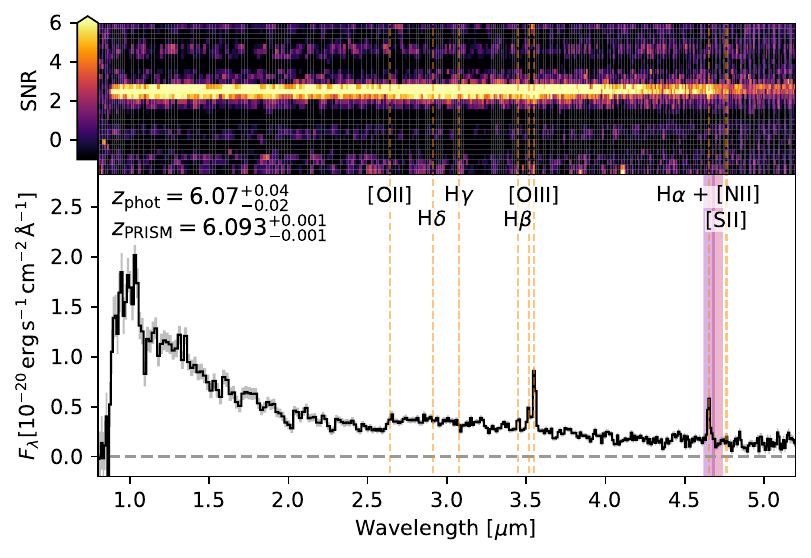}
    \includegraphics[width=0.98\columnwidth]{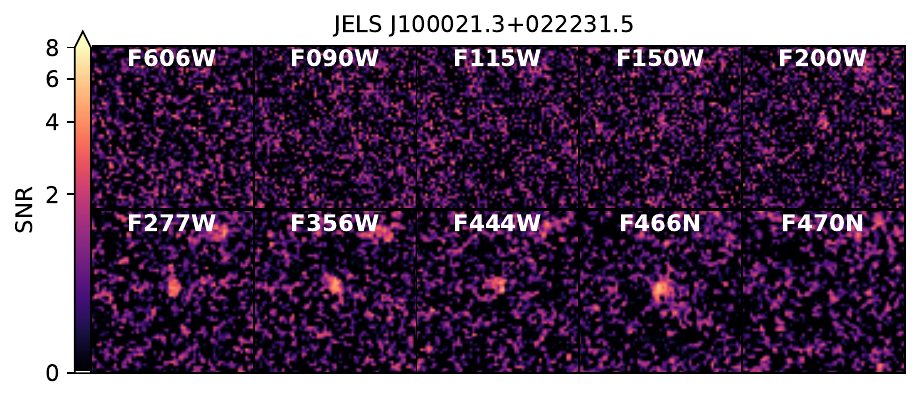}
    \includegraphics[width=0.98\columnwidth]{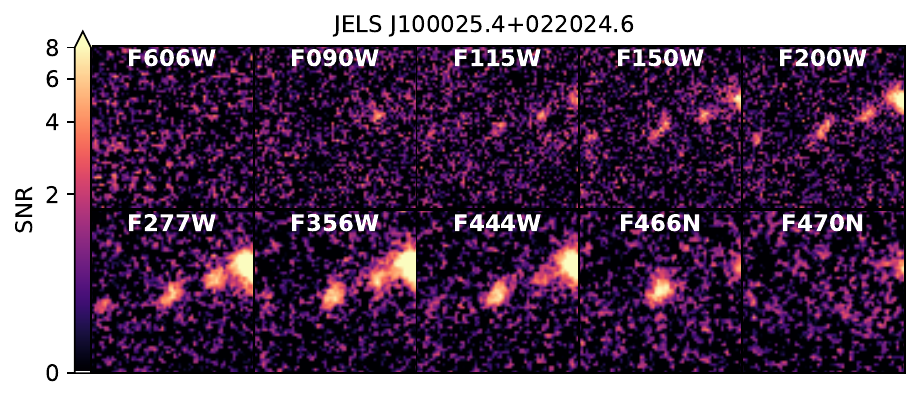}\vspace{-0.3cm}
    \includegraphics[width=0.98\columnwidth]{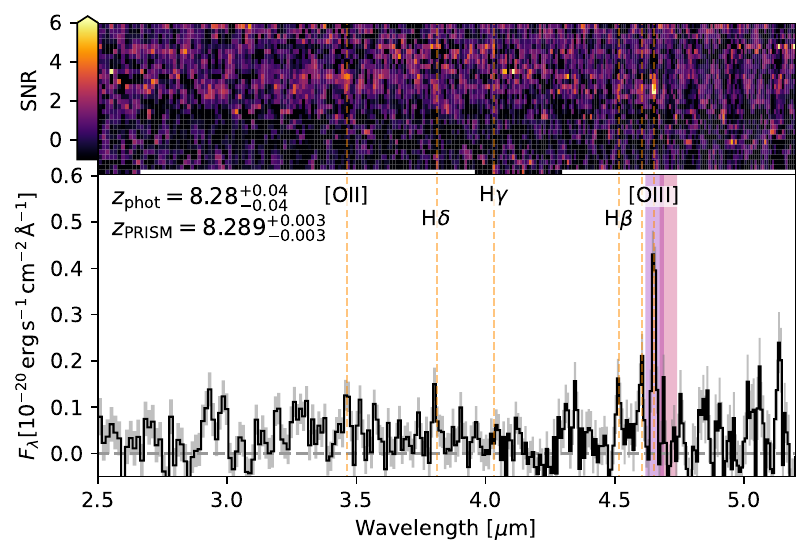}
    \includegraphics[width=0.98\columnwidth]{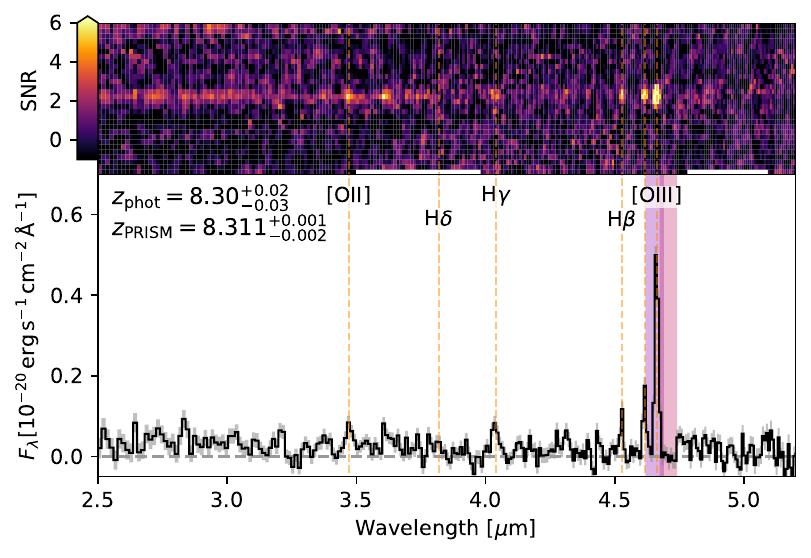}
    \caption{Example emission-line sources selected in JELS narrow-band imaging with spectroscopic confirmation from NIRSpec PRISM spectroscopy (DD 6585, PI Coulter). For each source, the upper panel shows photometry cutouts ($2 \,\textrm{arcsec} \times2 \,\textrm{arcsec}$) from HST/ACS (F606W) and JWST/NIRCam from both PRIMER and JELS. The lower panel for each source shows the corresponding standard pipeline reduced 2D and 1D PRISM spectra. A subset of bright rest-frame optical emission lines at the confirmed spectroscopic redshift are illustrated by the vertical dashed lines and corresponding labels. Shaded regions also show the wavelength coverage of the overlapping F466N and F470N filters for reference. The top row shows sources selected as \halpha emitters, while the bottom row shows sources selected as robust $z\sim8.3$ \oiii emitters. Note that at $z\sim8.3$, the width of the F466N/F470N narrow-bands is such that \mbox{[O\,{\sc iii]\sc{$\lambda$4960}}}\, and \mbox{[O\,{\sc iii]\sc{$\lambda$5008}}}\, (and potentially \hbeta) can be individually isolated from photometry alone.}
    \label{fig:examples}
\end{figure*}
% 11794: 5908.5        11272: 17725.5
% 10130: 11817.0       57066: 11817.0

In Fig.~\ref{fig:examples}, we show a subset of the available HST and JWST/NIRCam cutouts for each object as well as the standard MAST pipeline reduced 2D and 1D PRISM spectra.
Motivated by the expectation of high-EW emission-lines, we derive spectroscopic redshifts from the 1D PRISM spectra through a simple $\chi^{2}$-minimisation, fitting an emission-line template (convolved to the PRISM spectral resolution at 4.6\micron) to the continuum-subtracted spectrum over the wavelength range of $3 < \lambda < 5\micron$\footnote{Our derived spectroscopic redshifts are all in agreement with the independently derived estimates from the DAWN JWST Archive: \href{https://s3.amazonaws.com/msaexp-nirspec/extractions/nirspec_graded_v3.html}{https://s3.amazonaws.com/msaexp-nirspec/extractions/nirspec\_graded\_v3.html}.}.

The two confirmed \halpha sources are shown in the top row of Fig~\ref{fig:examples}, observed with exposure times of 5\,909 and 17\,725s respectively. 
JELS J100033.5+022355.5 (upper left panel) is representative of a large fraction of the \halpha emitters selected by JELS, with the overall spectral energy distribution (SED) dominated by high equivalent width (EW) emission lines.
The high-EW emission-line contribution can be inferred from the photometry cutouts alone, with \oiiia and \hbeta\, responsible for the F356W broadband showing clearly elevated flux; this is confirmed by the PRISM spectroscopy.
In comparison, JELS J100033.3+022331.7 (upper right panel) exhibits significant UV and optical continuum detections across the broadband SED, although F356W is still significantly enhanced relative to F277W.
The corresponding PRISM spectrum reveals significantly lower EW emission-lines, with evidence for a more evolved stellar population in the form of a clear Balmer break.
While both sources are correctly identified as $z\sim6$ sources when using only PRIMER observations, with the spectroscopic redshift contained within the photo-$z$ posterior, the widths of the photo-$z$ posteriors are significantly broader than the PRIMER+JELS estimates by up to $\sim10\times$ (consistent with the results from Section~\ref{sec:sims-photoz}).

The bottom row of Fig.~\ref{fig:examples} then presents the two \oiii-emitters selected from JELS, both with total on-source exposure times of 11\,817s.
The left-hand source, J100021.3+022231.5, is securely identified as a $z\sim8.3$ \oiii emitter by the photo-$z$ analysis to a precision of $<0.005\times(1+z)$, despite having only weak constraints on the Lyman break and very faint rest-UV continuum.
With $\text{SNR}_{\text{F115W}} < 1$ and $\text{SNR}_{\text{F150W}} < 4$ (in 0.3-arcsec diameter apertures), photo-$z$ estimates using broadband photometry only are limited to constraining the source as $z\gtrsim5$, but with a very broad posterior allowing solutions up to $z > 10$ (see Fig.~\ref{fig:pofz_nb_bb}).
Although faint, the extracted 1D PRISM spectrum for this source shows clear \hbeta, \oiiil, \oiii and \oii lines confirming the photo-$z$ redshift solution. 
The right-hand source, J100025.4+022024.6, is selected as a narrow-band excess by JELS, but is also bright enough in the rest-UV to be robustly selected as $z\sim8$ from broadband photometry with over 90 per cent of the PRIMER-only photo-$z$ posterior in the range of $8 < z < 9$ (and hence was also included in the DD filler programme from an independent selection).
We note that the rest-UV continuum is detected in the NIRSpec PRISM observation, however, we limit the wavelength range of the PRISM spectra presented in Fig.~\ref{fig:examples} to $>2.5$\micron\, to demonstrate the resolved \oiiil and \oiii lines.

Although only a limited sample, the confirmation of all four narrow-band excess selected sources and their diverse properties gives evidence that JELS offers a broad and robust selection in novel parameter space.
In addition to enabling unique science from the photometric data alone, the JELS emission-line samples are therefore ideal for future spectroscopic studies.

\section{Summary}\label{sec:summary}
We have presented an overview of the \emph{JWST} Emission Line Survey (JELS), a \emph{JWST} imaging survey designed to extend selection of emission-line galaxies using narrow-band filters into new redshift regimes using the JWST/NIRCam F466N and F470N filters.
Simultaneously, JELS aims to provide a window into the resolved properties of star-forming galaxies at cosmic noon with matching F212N and F200W observations. 
The JELS Cycle 1 observations presented cover $\sim63$ arcmin$^{2}$ within the wider PRIMER COSMOS legacy field (Dunlop et al., \emph{in prep.}), which provides both the key F444W broadband imaging necessary for F466N/F470N excess selection as well as extensive multi-wavelength imaging required for robust line identification.

We have demonstrated that the JELS imaging reaches the extremely high sensitivities required to achieve the survey's primary science goals.
Based on the distribution of $5\sigma$ limiting magnitudes in the F466N and F470N mosaics (in 0.3 arcsec diameter apertures), the limiting line fluxes are estimated to be $\sim1.2\times10^{-18}\,\text{erg s}^{-1}\text{cm}^{-2}$ over 90 per cent of the field, reaching up to $\sim2\times$ fainter emission lines than current slitless spectroscopic surveys in the literature.
For the primary science goal of probing \halpha at $z\sim6.1$, these flux limits correspond to $\log_{10}(L_{\text{H}\alpha}/ \text{erg\,s}^{-1}) \sim 41.53-41.76$, or unobscured star-formation rates of 0.9--1.3 $\text{M}_{\odot}\,\text{yr}^{-1}$.

The F212N narrow-band mosaic reaches line sensitivities of 1.4--2.5$\times10^{-18}\,\text{erg s}^{-1}\text{cm}^{-2}$, corresponding to $\log_{10}(L_{\text{H}\alpha}/ \text{erg\,s}^{-1}) \sim 40.75-40.99$; a factor of up to $\sim10$ further down the $z=2.23$ \halpha LF than previously available from ground-based narrow-band surveys.
The combination of this extraordinary depth with \emph{JWST}'s exquisite spatial resolution offers an unprecedented view of the resolved star-formation properties in galaxies at the peak of cosmic star formation history, for example by enabling detailed morphological comparison between \halpha, UV and in-situ stellar mass in representative samples of galaxies.

We have highlighted the unique science cases for the JELS observations: a census of \halpha emitters at $z\sim6.1$ that offers complementary constraints on the cosmic star-formation history and the galaxy population in the early Universe, novel probes of both early (through \oiiia emitters) and late stages (\halpha at $z\sim6.1$) of cosmic reionization, dust unbiased samples of star-forming galaxies at cosmic noon (\halpha/\Pa/\Pb), and spatially resolved properties of ionised gas in galaxies on sub-kpc scales at $2 \lesssim z \lesssim 6$.
Through detailed simulations of realistic \halpha and \oiii emitter populations, we have also demonstrated that for intrinsic line luminosities above the JELS limiting magnitudes, the resulting photo-$z$ estimates can be constrained to near spectroscopic accuracy ($\sigma_{\text{NMAD}} < 0.005\times(1+z)$) for a wide range of intrinsic equivalent widths.
These simulations also show that the emission line luminosities estimated from the JELS narrow-band excess can be both extremely accurate (bias less than $0.01$ dex) and measured with sufficient precision ($\pm0.05-0.1 \text{dex}$) that the limiting precision on intrinsic properties will the precision to which dust attenuation corrections can be made (as is also the case in slitless spectroscopic surveys).

Initial results for the primary \halpha sample at $z\sim6.1$ and full JELS-selected photometry catalogues are presented in a companion paper, \citet{Pirie2024}, with a number of further studies on the detailed properties of \halpha, \Pa/\Pb and \oiiia samples also in progress.
With JELS adding both novel wavelength information (F212N, F466N, F470N) and significant additional broadband sensitivity (F200W) within one of the key extra-galactic legacy fields, we expect the broader scientific return from the community to extend far beyond these initial goals.

\section*{Acknowledgements}
We thank the anonymous referee for their helpful and constructive feedback that has significantly improved this manuscript.
The authors also thank David Coulter and Armin Rest for allowing the inclusion of JELS targets in their director's discretionary observing programme.
KJD acknowledges support from the Science and Technology Facilities Council (STFC) through an Ernest Rutherford Fellowship (grant number ST/W003120/1). DJM, PNB, RK and RJM acknowledge the support of the UK STFC via grant ST/V000594/1.
PNB and RK are grateful for support from the UK STFC via grant ST/Y000951/1.
RKC was funded by support for program \#02321, provided by NASA through a grant from the Space Telescope Science Institute, which is operated by the Association of Universities for Research in Astronomy, Inc., under NASA contract NAS5-03127. RKC is grateful for support from the Leverhulme
Trust via the Leverhulme Early Career Fellowship.
JSD acknowledges the support of the Royal Society via a Royal Society Research Professorship.
CLH acknowledges support from the Leverhulme Trust through an Early Career Research Fellowship and also acknowledge support from the Oxford Hintze Centre for Astrophysical Surveys which is funded through generous support from the Hintze Family Charitable Foundation. 
E.I. gratefully acknowledge financial support from ANID - MILENIO - NCN2024\_112 and ANID FONDECYT Regular 1221846.
AMS and IRS acknowledge support from the STFC via grant ST/X001075/1.
This work was initiated in part at Aspen Center for Physics, which is supported by National Science Foundation grant PHY-2210452.
For the purpose of open access, the author has applied a Creative Commons Attribution (CC BY) licence to any Author Accepted Manuscript version arising from this submission.

%%%%%%%%%%%%%%%%%%%%%%%%%%%%%%%%%%%%%%%%%%%%%%%%%%
\section*{Data Availability}

The data underlying this article are available in the Mikulski Archives for Space Telescopes (MAST) Portal under proposal ID numbers 2321 (JELS imaging) and 6585 (NIRSpec PRISM spectroscopy). Higher level data products, including all reduced mosaics in the JELS narrow and broadband filters (v0.8 and v1.0), as well as associated catalogues presented in \citet{Pirie2024} are made available through the \href{https://datashare.ed.ac.uk}{Edinburgh DataShare} service. Any other data produced for the article will be shared on reasonable request to the corresponding author.

%%%%%%%%%%%%%%%%%%%% REFERENCES %%%%%%%%%%%%%%%%%%

% The best way to enter references is to use BibTeX:

\bibliographystyle{mnras}
\bibliography{jels_survey_paper} % if your bibtex file is called example.bib

% Alternatively you could enter them by hand, like this:
% This method is tedious and prone to error if you have lots of references
%\begin{thebibliography}{99}
%\bibitem[\protect\citeauthoryear{Author}{2012}]{Author2012}
%Author A.~N., 2013, Journal of Improbable Astronomy, 1, 1
%\bibitem[\protect\citeauthoryear{Others}{2013}]{Others2013}
%Others S., 2012, Journal of Interesting Stuff, 17, 198
%\end{thebibliography}

%%%%%%%%%%%%%%%%%%%%%%%%%%%%%%%%%%%%%%%%%%%%%%%%%%

%%%%%%%%%%%%%%%%% APPENDICES %%%%%%%%%%%%%%%%%%%%%

\appendix
\section{JELS narrow-band mosaic release versions}\label{sec:app-versions}
As outlined in Section~\ref{sec:overview}, the initial JELS observations were impacted by severe scattered light.
Initial versions of the JELS imaging products, including those used in \citet[][v0.8]{Pirie2024}, made use of the data available and analysis pipelines at the time of analysis.
The imaging products and associated depths presented in this manuscript correspond to the versions with all JELS observing programme (GO 2321) data acquired (v1.0).
For completeness, here we outline the key differences in the JELS mosaic release versions and provide a quantitative comparison of the limiting emission-line sensitivities achieved by the respective versions.
The key data and pipeline differences between images version are:
\begin{itemize}
    \item v0.8: Image reduction using PENCIL version based on \emph{JWST} pipeline version 1.10.2 (\texttt{jwst\_1107.pmap}). Mosaics incorporate all frames observed May 2023, with scattered light contribution subtracted from LW filters (F466N/F470N) and scattered light masked from impacted SW frames (F212N/F200W).
    \item v1.0: Image reduction using PENCIL version based on \emph{JWST} pipeline version 1.13.4 (\texttt{jwst\_1303.pmap}). Mosaics incorporate all frames observed May 2023 and November 2024. Scattered light contributions subtracted from LW (F466N/F470N) and masked in SW (F212N/F200W) frames from May 2023, with corresponding November 2024 frames included in addition.
\end{itemize}
\begin{figure}
    \centering
    \includegraphics[width=0.98\columnwidth]{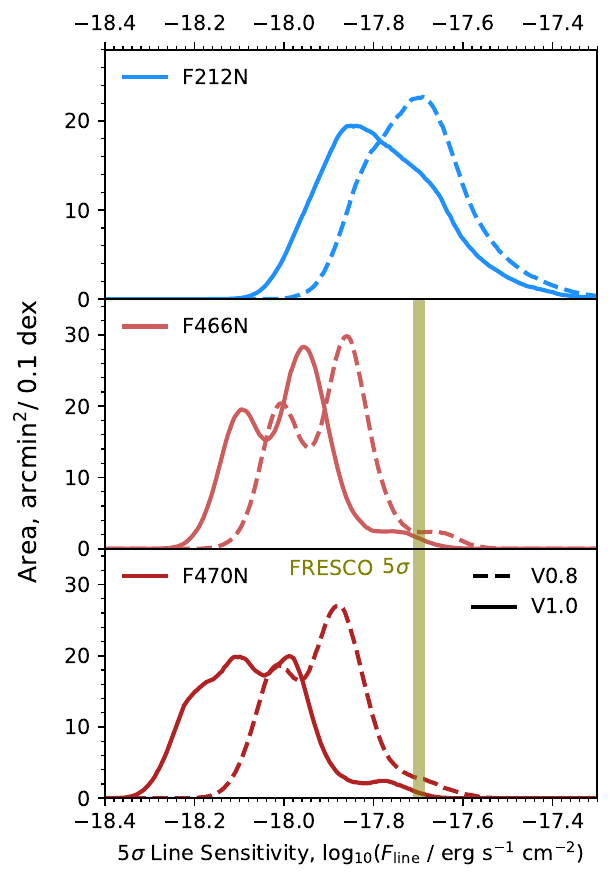}
    \caption{Emission line sensitivity of the JELS narrow-bands as a function of area per limiting sensitivity for v0.8 of the JELS mosaics \citep[e.g.][dashed lines]{Pirie2024} and the v1.0 mosaics presented here (solid lines). Also plotted for reference for the F466N and F470N panels is the average 5$\sigma$ limiting line sensitivity for the FRESCO slitless spectroscopy survey \citep[thick vertical line;][]{Oesch2023}.}
    \label{fig:sensitivity_changes}
\end{figure}

In Fig.~\ref{fig:sensitivity_changes}, we present the area per limiting sensitivity in each narrow-band filter for both v0.8 and v1.0 of the JELS mosaics following the procedure outlined in Section~\ref{sec:properties} (cf. Fig.~\ref{fig:sensitivity_curves}).
For F466N, where the scattered light impact was negligible and the input imaging for v0.8 and v1.0 is effectively unchanged, we find that the v1.0 mosaic achieves a consistent $\sim0.09\, \text{dex}$ improvement in sensitivity across the full image; indicative of general NIRCam sensitivity improvements from revisions to the relevant read-noise and flat-field calibration files from the later CRDS version.
For F470N, the sensitivity distribution includes the same $\sim0.09\,\text{dex}$ systematic shift as for F466N, but with an additional increase in sensitivity from the repeat observations; a higher fraction of the mosaic is covered by two full visits at the full JELS exposure time, with an additional fraction now observed at $3\times \sim6000\text{s}$.

For F212N, which was most severely impacted by the enhanced scattered light, the peak of the sensitivity distribution is $\sim0.2\,\text{dex}$ fainter, reflecting the significantly increased area with $2\times \sim6000\text{s}$ visits (per the original survey design), a subset of area now with $3\times$ visits due to repeats, and similar systematic gains from calibration improvements.
Based on the change in sensitivity of the shallowest regions of the field where no new data is included, we estimate that the systematic improvement in F212N sensitivity from CRDS reference file changes is smaller than for the LW bands, at the level of $\sim0.05\,\text{dex}$.

For all three JELS narrow-band filters, we note that while the gains in sensitivity for v1.0 are consistent and statistically significant with respect to our ability to constrain the image noise itself, the changes are not as scientifically significant.
Impacts to the robust emission line samples produced from JELS imaging and the predictions of the simulations presented in Section~\ref{sec:sims} are negligible between versions.
Sample sizes from v1.0 may be increased in size, but only at the 10-20\% level.
However, the inferred properties and redshifts for individual objects are as robust in v0.8 as in v1.0 due to the more significant impacts of large photometric uncertainties in other filters for sources near the detection limit and the standard inclusion of 5-10\% flux uncertainties for photo-$z$/SED fitting analysis.

\section{Correcting for \oiiil contributions in \oiii luminosity estimation}\label{sec:app-oiii_corr}
For the \oiii emission line selection, the JELS narrow-bands are sufficiently narrow that typically only one of the \oiiia doublet lines contributes the majority of flux at any given redshift \citep[unlike for some lower redshift narrow-band surveys where both \oiiia line, and sometimes \hbeta\,, are often encompassed by the narrow-band filter, e.g.][]{khostovan2016}.
The contribution from the secondary line, i.e. \oiiil for \oiii excess selection, however, is non-negligible and naive estimates of the line flux based on the narrow-band colour excess alone could therefore over-estimate the true \oiii line flux.
Furthermore, given the high-EWs of the $z\sim8.3$ samples, the \oiiil emission will have significant contribution to the surrounding broadband flux, $f_{\lambda,\rm{F444W}}$, that could lead to an overestimate of the stellar continuum level and hence an underestimate of the inferred \oiii flux. 
Formally, high-EW \hbeta\, emission will also contribute to an overestimate of the stellar continuum, however given the high \oiiia/\hbeta\, ratios observed at $z > 6$ \citep{shapley2023a,sanders2023}, we assume that any resulting corrections would be negligible relative to the photometric uncertainties.
Regardless, the relative balance of these two competing secondary or tertiary line contaminations will depend on the precise redshift. 
We therefore implement a simple analytic correction based on the expected relative contribution of the \oiiia lines to both the narrow-band and broad-band filter fluxes.
For the simplifying assumption that the emission lines themselves have effectively no velocity dispersion, and assuming the standard ratio of \oiii/\oiiil$=2.98$, the relative contributions of both \oiiia lines to a given filter, $f_{\textsc{Oiii},i}$, is proportional to
\begin{equation}
    \begin{split}
   f_{\textsc{Oiii},i} \propto&\, f_{\lambda5008}\tilde{t}_{i,\lambda5008}(z) + \frac{f_{\lambda5008}}{2.98}\tilde{t}_{i,\lambda4960}(z), \quad \text{or} \\
   f_{\textsc{Oiii},i} \propto&\, f_{\lambda5008}\left (  \tilde{t}_{i,\lambda5008}(z) +  \frac{\tilde{t}_{i,\lambda4960}(z)}{2.98} \right )
   \end{split}
\end{equation}
where $\tilde{t}_{i,\text{line}}(z)$ is the filter throughput at the wavelength of the corresponding line at redshift $z$, normalised to the peak filter throughput.
The redshift dependent correction factor, $C_{i}(z)$, to relate the observed flux to that corresponding to only the \oiii flux can then simply be defined as
\begin{equation}
    C_{i}(z) = \frac{1}{\tilde{t}_{i,\lambda5008}(z) + \frac{\tilde{t}_{i,\lambda4960}(z)}{2.98}}.
\end{equation}
Eq.~\ref{eq:f_line_1} can then be modified such that the \oiii emission line flux can estimated as
\begin{multline}\label{eq:f_line_2}
F_{\lambda 5008}(z) \ = \ \Delta \lambda_{\rm{F466N}} \ \frac{f_{\lambda,\rm{F466N}}C_{\rm{F466N}}(z) \ - \ f_{\lambda,\rm{F444W}}C_{\rm{F444W}}(z)}{1 \ - \ (\Delta \lambda_{\rm{F466N}} / \Delta \lambda_{\rm{F444W}})}\\
\text{erg s}^{-1} \text{cm}^{-2}.
\end{multline}
From tests convolving model \oiiia emission lines with the F466N and F444W filters over a range of plausible velocity dispersions and across the redshift range probed by the excess selection, we find that the simplifying assumption of infinitely narrow intrinsic lines is accurate to $\sim1\%$ for intrinsic velocity dispersions with FWHM less than 250 $\text{km s}^{-1}$.
When estimating the line luminosity for an \oiii source, we then draw 100 redshifts from the photo-$z$ posterior, calculating $F_{\lambda 5008}(z)$ using Eq.~\ref{eq:f_line_2} and incorporating the redshift into the corresponding luminosity distance calculation.
The estimated line luminosity for an individual source including the uncertainty from the redshift dependent \oiiil contribution can then be derived from the resulting distribution (i.e. 16, 50 and 84th percentiles).

% If you want to present additional material which would interrupt the flow of the main paper,
% it can be placed in an Appendix which appears after the list of references.

%%%%%%%%%%%%%%%%%%%%%%%%%%%%%%%%%%%%%%%%%%%%%%%%%%

% Don't change these lines
\bsp	% typesetting comment
\label{lastpage}
\end{document}